\RequirePackage{fix-cm}
\documentclass[smallextended]{svjour3}       % onecolumn (second format)

\smartqed  % flush right qed marks, e.g. at end of proof
\usepackage{tabu} % only used for the table example
\usepackage{booktabs} % only used for the table example
\usepackage{comment}
\usepackage{graphicx}
\usepackage{paralist}
\usepackage{xspace}
\usepackage{natbib}
\usepackage{caption}
\usepackage{subcaption}
\usepackage[pdftex,colorlinks=true,pdfstartview=FitV,
 	linkcolor=black,citecolor=black,urlcolor=black,bookmarks=false]{hyperref}
\usepackage{mathptmx}      % use Times fonts if available on your TeX system
\usepackage{latexsym}
\newcommand{\ie}{\emph{i.e.},\xspace}
\newcommand{\eg}{\emph{e.g.},\xspace}
\newcommand{\etal}{\emph{et al.}\xspace}
\usepackage{paralist}
\usepackage{geometry}
\usepackage{amssymb}
\usepackage{ifthen}
\usepackage[normalem]{ulem} % for \sout
\usepackage{xcolor}

\newboolean{showedits}
\setboolean{showedits}{true} % toggle to show or hide edits
\ifthenelse{\boolean{showedits}}
{
  \newcommand{\mehf}[1]{\textcolor{green}{\uwave{#1}}} % please rephrase
  \newcommand{\del}[1]{\textcolor{red}{\sout{#1}}} % please delete
}{
  \newcommand{\del}[1]{} % please delete
  
}
\newboolean{showcomments}
\setboolean{showcomments}{true}
\newcommand{\id}[1]{$-$Id: scg-llncs.tex 30911 2010-02-05 10:21:47Z oscar $-$}

\ifthenelse{\boolean{showcomments}}
{\newcommand{\nbc}[3]{
 {\colorbox{#3}{\bfseries\sffamily\scriptsize\textcolor{white}{#1}}}
 {\textcolor{#3}{\sf\small$\blacktriangleright$\textit{#2}$\blacktriangleleft$}}}
 }
{\newcommand{\nbc}[3]{}
 \renewcommand{\del}[1]{} % please delete
 }

\newcommand\lm[1]{\nbc{LM}{#1}{blue}}

\usepackage{xspace}
\usepackage{paralist}
\journalname{Journal of Visualization}
\begin{document}

\title{A Systematic Literature Review of Modern Software Visualization
}


\author{Noptanit Chotisarn \and
         Leonel Merino \and
         Xu Zheng \and
         Supaporn Lonapalawong \and
         Tianye Zhang \and
         Mingliang Xu \and
         Wei Chen %etc.
}


\institute{Noptanit Chotisarn, Supaporn Lonapalawong, Tianye Zhang, Wei Chen \at
              State Key Lab of CAD\&CG, Zhejiang University, Hangzhou, China \\
              \email{chotisarn@zju.edu.cn, 11821132@zju.edu.cn, zhangtianye1026@zju.edu.cn,  chenwei@cad.zju.edu.cn} \\
              Wei Chen is the corresponding author.%  \\
%             \emph{Present address:} of F. Author  %  if needed\textbf{}
           \and
           Leonel Merino \at
              Visualization Research Center, University of Stuttgart, Stuttgart, Germany \\
              \email{leonel.merino@visus.uni-stuttgart.de}
           \and   
           Xu Zheng \at
              School of Computer Science, Zhejiang University, Hangzhou, China \\
              \email{xz15968526011@hotmail.com}
            \and
            Mingliang Xu \at
            College of Computer Science, Zhengzhou University, Zhengzhou, China \\
            \email{iexumingliang@zzu.edu.cn}
}

\date{Received: date / Accepted: date}

\maketitle
\begin{abstract}
We report on the state-of-the-art of software visualization. To ensure reproducibility, we adopted the Systematic Literature Review methodology. That is, we analyzed 1440 entries from IEEE Xplore and ACM Digital Library databases. We selected 105 relevant full papers published in 2013--2019, which we classified based on the aspect of the software system that is supported (\ie structure, behavior, and evolution). For each paper, we extracted main dimensions that characterize software visualizations, such as software engineering tasks, roles of users, information visualization techniques, and media used to display visualizations. We provide researchers in the field an overview of the state-of-the-art in software visualization and highlight research opportunities. We also help developers to identify suitable visualizations for their particular context by matching software visualizations to development concerns and concrete details to obtain available visualization tools.

\keywords{Software visualization \and Systematic literature review \and Information visualization}
\end{abstract}

%%%%%%%%%%%%%%%%%%%%%% INTRODUCTION %%%%%%%%%%%%%%%%%%%%%%
\section{Introduction}
\label{intro}
\maketitle
The visualization of data in multiple domains has shown to be effective in supporting users to answer complex and frequent questions~\citep{liu2014survey}. Software engineering is a domain that has been benefited from visualization~\citep{diehl2007software}. In \emph{software visualization}, users are usually developers, and their tasks represent development concerns. Indeed, many software visualization approaches and systems have been proposed during the last decades. These visualizations can be classified into one of three categories depending on the visualized aspect of the subject software system~\citep{diehl2007software}: 
\begin{inparaenum}[(\itshape i\upshape)]
  \item \emph{structure}, \eg to analyze how the structure of the source code of mobile applications differs from traditional software systems~\citep{minelli2013software}, to ease code reading~\citep{zhu2019people};
  \item \emph{behavior}, \eg to optimize the performance of large-scale parallel programs based on execution traces~\citep{isaacs2014combing}, to manage cloud computing systems~\citep{xu2019clouddet}; and 
  \item \emph{evolution}, \eg to analyze a fine-grained code change history~\citep{yoon2013visualization}, to explore code quality~\citep{mumtaz2019exploranative}.
\end{inparaenum}
Despite the many proposed software visualizations that have proven to be effective to support developers in software engineering tasks, there is limited adoption of software visualizations by developers, for instance, in context of corrective maintenance and debugging ~\citep{sensalire2008classifying}. We hypothesize that visualization approaches proposed to support software engineering tasks might include particular dimensions that differ with the state-of-the-art in information visualization. Consequently, our interest is to investigate not only software concerns tackled by visualizations, but also employed visualization techniques, roles of users, and media used to display visualizations. 

The goal of our investigation is twofold:

\begin{enumerate}[(\itshape i\upshape)]
  \item to provide researchers in the field an overview of the state-of-the-art in software visualization, and highlight research opportunities; and
  \item to help developers to identify suitable visualizations for their particular context by matching software visualizations to development concerns.
\end{enumerate}

To achieve our goal, we conducted a systematic literature review of software visualization. %\del{We consider both, software visualizations introduced in the literature recently (\ie during the past six years) and also older techniques which have been proven to be relevant.} 
For each proposed visualization, we identify software visualization aspects, software engineering tasks, roles of  expected users, mediums used to display visualizations, employed information visualization techniques, and name's of prototypical tools.

%tasks
We found that most software visualizations support tasks that involve design and implementation, and maintenance of software systems. Though we did not find visualizations proposed to support software requirements, undoubtedly, we think they exist, and they might not appeared in our results because the our search process.
%vis technique
We observed that software visualizations frequently use multivariate visualization techniques. Amongst them, empirical methods is the most frequent technique, which is used to assist in the design of effective software visualizations.
%medium
The majority of software visualizations are visualized using standard computer screens; however, we observe that several approaches that support the analysis of the structure of systems are displayed in immersive 3D environments (\eg virtual reality). Interestingly, we found that there is an increasing number of software visualizations that are displayed in a medium different than the computer screen.

The remainder of the paper is structured as follows:  Section~\ref{sec:related} elaborates on related work;  Section~\ref{sec:method} describes the methodology that we followed to collect and select relevant studies in software visualization; Section~\ref{sec:results} presents our results by classifying evaluations based on the adopted categories; Section~\ref{sec:discussion} discusses the results in terms of our goals and limitations of our findings; and Section~\ref{sec:conclusion} concludes and presents future work.

%%%%%%%%%%%%%%%%%%%%%% RELATED WORK %%%%%%%%%%%%%%%%%%%%%%
\section{Related work}
\label{sec:related}
Several studies that have surveyed the literature in information visualization have focused on how information and knowledge transform into interactive visual representations across multiple domains~\citep{geisler1998making,yi2007toward,rodrigues2015survey,kumar2016review,mcnabb2017survey}. 
Various reviews of the literature in information visualization have classified the state-of-the-art research and applications in multiple ways. For instance, a survey~\citep{sun2013survey} analyzed visual analytics that provides an analytic space, including space and time, multivariate, text, graph and network, and other, such as applications cross-visual mapping, model-based analysis, and user interactions. Moreover, in a specific field, such as machine learning (ML), it is also represented from a visual analytics perspective that classifies the relevant work in ML field into three categories: understanding, diagnosis, and refinement~\citep{liu2017towards}. We observe there is a study~\citep{liu2014survey}, which we consider particularly useful to classify proposed visualization approaches. Consequently, we adopted this classification in our study to classify software visualizations. In it, visualization approaches are classified into one of four main categories: \emph{Empirical methodologies}, \emph{Interactions}, \emph{Frameworks}, and \emph{Applications}.

Several other studies have mainly focused on software visualization. A decade ago, Storey \etal~\citep{storey2005use} proposed a framework to classify software visualizations based on human aspects of software engineering (\ie intent, information, presentation, interaction, and effectiveness). 
Kienle and Muller~\citep{kienle2007requirements} identified requirements to build and evaluate tools in software visualization. They focused on visualizations for software maintenance, re-engineering, and reverse engineering. 
We consider these studies important for the software visualization field. In consequence, we include in our study the aspects that these studies have analyzed.
More recently, Shahin \etal~\citep{shahin2014systematic} investigated software visualizations that support understanding software architectures of large and complex systems, and classified visualization techniques used to support software architectures. Beck \etal~\citep{beck2017taxonomy} conducted a survey on dynamic graph visualization, and analyzed the use of dynamic graphs to visualize  software systems. They found that dynamic graph approaches have been customized to various application scenarios, one of them is software engineering.
Mattila \etal~\citep{mattila2016software} presented an overview of recent software visualizations introduced in the literature. They found that only a few software visualization papers address software processes. Greene \etal~\citep{greene2017visualizing} proposed a tag could-based visualization for the analysis of software project repositories. The tags visualization example was mentioned in the guest editor of the special section on software visualization by Bergel \etal~\citep{bergel2017guest}. 
In our study, we do not only analyze software processes, but we also analyze visualization techniques, and media used to display visualizations. Merino \etal~\citep{merino2016towards,merino2018towards} studied fairly similar development concerns.  However, there are two fundamental differences to our investigation: 1) inclusion criteria, and 2) classification method. That is, they included papers from only two venues (\ie SOFTVIS and VISSOFT) and published until 2016, and employed a classification of visualization techniques that we think is less applicable to software visualization as the one we chose.    
Maletic \etal~\citep{maletic2002task} elaborates on a discussion concerning a taxonomy of software visualization related.

% To classify papers by software processes based on a set of related four process activities~\citep{sommerville2011software} that are fundamental to software engineering:
% \begin{inparaenum}[(\itshape i\upshape)]
% \item \emph{software specification} (specifically software requirements), which deals with the specification of software functionality and its constraints definition process.
% \item \emph{software design and implementation}, which translates a specification into a concrete software product.
% \item \emph{software validation} or \emph{software testing}, which prevents that a software system exhibits incorrect behavior.
% \item \emph{software evolution} (specifically software maintenance), which deals with changes to a software system to deal with new requirements and discovered bugs that need to be fixed.
% \end{inparaenum}

%%%%%%%%%%%%%%%%%%%%%% LITERATURE REVIEW METHODS %%%%%%%%%%%%%%%%%%%%%%
\section{Methodology}
\label{sec:method}
\begin{figure}[htbp]
 \centering 
 \includegraphics[width=\linewidth]{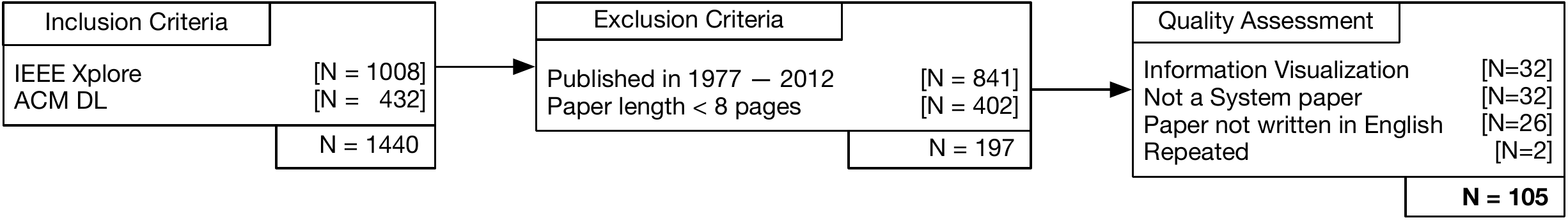}
 \caption{Stages of the search process and number of selected studies at each stage.}
  \label{fig:included_papers_graph}
\end{figure}

We apply the systematic literature review (SLR) approach that ensures a rigorous research methodology for evidence-based software engineering. We follow Kitchenham's guidelines~\citep{kitchenham2000guidelines} to mitigate bias in the results of the literature survey. We select this method because it offers a means for evaluating and interpreting relevant research to a topic of interest by evidence, which is robust and transferable.
We apply the method by defining a review protocol that ensures rigor and reproducibility. We determine
\begin{inparaenum}[(\itshape i\upshape)]
  \item data sources and search strategy,
  \item inclusion and exclusion criteria,
  \item quality assessment,
  \item data extraction, and
  \item selected studies.
\end{inparaenum}
An overview of the selection process is shown in Figure~\ref{fig:included_papers_graph}.

\subsection{Data sources and search strategy}
Amongst popular digital libraries and search engines that are available to collect papers, we select the ACM Digital Library\footnote{\url{http://dl.acm.org/}} and the IEEE Xplore\footnote{\url{http://ieeexplore.ieee.org}}. We select them because they have indexed the proceedings of the VISSOFT conference, which is especially dedicated to software visualization. Moreover, they also contain relevant papers from higher ranked venues, \eg TVCG, ICPC, ICSE, VINCI, ASWEC, ICSME.

\subsection{Inclusion and exclusion criteria}
We define a search query for retrieving relevant papers in software visualization as follows: 

\vspace{0.5em}
{\centering
\noindent\framebox{
    \parbox[t][0.5cm]{.8\linewidth}{
    \addvspace{0.1cm} \centering 
      \emph{``software visualization'' OR ``software visualisation''}
    } 
  }\\
\vspace{0.5em}}

We execute the query in ACM Digital Library and obtained 432 records\footnote{Last visit, Mar. 27, 2019}. Similarly, we execute the query in IEEE Xplore and obtained 1008 records. We then exclude papers that might not be relevant in the scope of our investigation. In particular, from the 1440 records returned by the searched digital libraries, we exclude 1243 papers that corresponded to:
\begin{inparaenum}[(\itshape i\upshape)]  
  \item 841 papers published between 1977 and 2012, and %343 +498 +1
  \item 402 papers of less than 8 pages. %40 +319 +42
\end{inparaenum}

\subsection{Quality assessment}
We then assess the quality of the remaining 161 papers. First, we identify 26 papers that are not written in English, which we exclude. Although we specify keywords that target specifically software visualization, we observe that some papers returned by the engines relate to visualizations proposed to support concerns in other domains. Consequently, we exclude 32 of such papers, which relate to information visualization. Next, to identified relevant software visualization papers that describe systems (from which we can extract characteristics of employed visualization techniques and the software concern that is supported), we classify the studies according to the categories proposed by Munzner~\citep{munzner2008process}. In it, a visualization paper can be classified into one of five categories: system, technique, evaluation, design study, and model. 

We select system papers as they focus on architectural choices made in the design of an infrastructure, a framework, or a toolkit, as well as, focus on lessons learned from building a system and observing its use. In the rest of the paper, we simply call them as system papers.

For each paper, we read a title, an abstract, and a conclusion. In the cases where we still were not sure of the contributions, we scan through the sections of the rest of the paper. Although papers usually exhibit characteristics of more than one type, we classify the papers by focusing on their primary contribution. Consequently, we exclude 32 papers that do not describe systems (\eg literature reviews). When we find papers that address similar research questions that ours, for example, the work of Sensalire \etal~\citep{sensalire2008classifying}, we discuss them into the related work.%s and put into the knowledge tank for research.

\subsection{Data extraction}
We extract six items from each selected paper. Those are:
\begin{enumerate}[(\itshape i\upshape)]
  \item Software aspects: we extract aspects of the software system that is visualized: structure, behavior, and evolution.
  \item Software processes: we extract processes based on five categories: software requirements, software design and implementation, software validation, software maintenance, and all software processes.
  \item Software engineering roles: we extract roles played by stakeholders in software engineering, who are the target users of software visualizations.
  \item Information visualization techniques: we extract visualization techniques based on four categories:empirical methodologies, interaction, systems and frameworks, and applications.
  \item Display mediums: we extract the medium used to render a software visualization.
  \item Tools: we extract a tool's name, and URL.
\end{enumerate}

\subsection{Selected studies}
\label{sec:included}
We include in our study 105 papers, of which, 57 papers have been published in VISSOFT. The remaining 48 papers have been published across 38 other venues. In Table~\ref{tab:papers_venues}, we present the list of the venues, rank, and number of papers. We searched the rankings of involved conferences and journals using two sources: CORE Rankings Portal\footnote{http://www.core.edu.au/conference-portal} and Conference Ranks\footnote{http://www.conferenceranks.com/}. In the former, we used the ``Excellence in Research in Australia'' (ERA) database (\ie A [highest] to C [lowest]), and in the latter, we used the Qualis database (\ie A1 [highest] to B5 [lowest]).%\lm{for each add in parenthesis the scale used in the rank, e.g., A* excellent, A good, B...}%, which is a conference ranking has been published by the Brazilian ministry of education and uses the H-index as a performance measure for conferences.}
\begin{comment}
\begin{figure}[htbp]
 \centering 
 %%\includegraphics[width=0.9\linewidth]{pictures/chinavis_years.pdf}
 \includegraphics[width=0.9\linewidth]{pictures/newpictures/2013-2018.png}
 \caption{A bar chart of the distribution of 88 papers by year and VISSOFT compare with other venues.}
  \label{fig:papers_years}
\end{figure}
\end{comment}

%%%%%%%%%%%%%%%%%%%%%% RESULTS %%%%%%%%%%%%%%%%%%%%%%
\section{Results}
\label{sec:results}

\begin{comment}
\begin{figure*}[htbp]
 \centering 
 %\includegraphics[width=0.9\linewidth]{pictures/chinavis_venues.pdf}
 %\includegraphics[width=0.95\linewidth]{pictures/newpictures/venues.png}
 \includegraphics[width=0.95\linewidth]{pictures/WordItOut-word-cloud-4055140.png}
 \caption{A treemap of the distribution of papers by venues. \mehf{A word cloud represents the distribution of papers by venues.}\lm{this figure is not very interesting. I think any chart will not be interesting as most papers come from vissoft and only a few come from other venues. Instead you can create a table with 1) the acronym, 2) the full name (standardized) of the venues, 3) the ranking you can have a look at CORE, and 4) the number of papers found. As there are several venues that information could be somehow interesting.}}
  \label{fig:papers_venues}
\end{figure*}
\end{comment}

%\newcolumntype{R}{>{\raggedright\arraybackslash}m{7cm}}
\begin{table}[!ht]%[htbp]
  \caption{The list of distribution of papers by venues.}
  \label{tab:papers_venues}
  \setlength\tabcolsep{2pt}
 % \scriptsize%
  \centering%
  \begin{tabu}{llllc}
  \toprule
   \textbf{Abbrv.} &\textbf{Name} &\textbf{Rank} &\textbf{Source} &\textbf{\#}\\
  \midrule
\href{	http://vissoft-conferences.dcc.uchile.cl/	}{	VISSOFT	}	&	IEEE International Working Conference on Software Visualisation	&	B	&	ERA	&	57	\\
\href{	https://www.computer.org/csdl/journal/tg	}{	TVCG	}	&	IEEE Transactions on Visualization and Computer Graphics	&	A	&	ERA	&	4	\\
\href{	https://www.program-comprehension.org/	}{	ICPC	}	&	IEEE International Conference on Program Comprehension	&	C	&	ERA	&	3	\\
\href{	http://www.icse-conferences.org/	}{	ICSE	}	&	International Conference on Software Engineering	&	A1	&	Qualis	&	3	\\
\href{	https://vinci-conf.org/	}{	VINCI	}	&	Visual Information Communications International	&	C	&	ERA	&	3	\\
\href{	https://ieeexplore.ieee.org/xpl/conhome/1000682/all-proceedings	}{	ASWEC	}	&	Australian Software Engineering Conference	&	B	&	ERA	&	2	\\
\href{	https://conferences.computer.org/icsm/	}{	ICSME	}	&	IEEE International Conference on Software Maintenance and Evolution	&	A	&	ERA	&	2	\\
\href{	https://link.springer.com/conference/pgcic	}{	3PGCIC	}	&	International Conference on P2P, Parallel, Grid, Cloud and Internet Computing	&	B4	&	Qualis	&	1	\\
\href{	https://www.aeroconf.org/	}{	AeroConf	}	&	IEEE International Aerospace Conference	&	N/A	&	N/A	&	1	\\
\href{	http://www.nise.org/APSEC/	}{	APSEC	}	&	Asia-Pacific Software Engineering Conference	&	B	&	ERA	&	1	\\
\href{	http://ase-conferences.org/	}{	ASE	}	&	Automated Software Engineering Conference	&	A	&	ERA	&	1	\\
\href{	https://www-01.ibm.com/ibm/cas/cascon/	}{	CASCON	}	&	Annual International Conference on Computer Science and Software Engineering	&	B1	&	Qualis	&	1	\\
\href{	https://conferences.computer.org/chase2019/	}{	CHASE	}	&	International conference on Connected Health: Applications, Systems and Engineering Technologies	&	N/A	&	N/A	&	1	\\
\href{	http://www.swinflow.org/confs/2019/cit/	}{	CIT	}	&	IEEE International Conference on Computer and Information Technology	&	C	&	ERA	&	1	\\
\href{	https://2019.ecoop.org/home/COP-2019	}{	COP	}	&	International Workshop on Context-Oriented Programming	&	N/A	&	N/A	&	1	\\
\href{	https://dblp.org/db/conf/csmr/index	}{	CSMR	}	&	European Conference on Software Maintenance and Reengineering	&	C	&	ERA	&	1	\\
\href{	https://edoc2019.sciencesconf.org/resource/page/id/7	}{	EDOC	}	&	IEEE International Enterprise Distributed Object Computing Conference	&	B	&	ERA	&	1	\\
\href{	http://www.enase.org/	}{	ENASE	}	&	International Conference on Evaluation of Novel Approaches to Software Engineering	&	B	&	ERA	&	1	\\
\href{	http://www.foundationsofdigitalgames.org/	}{	FDG	}	&	International Conference on the Foundations of Digital Games	&	C	&	ERA	&	1	\\
\href{	https://gecco-2020.sigevo.org/index.html/HomePage	}{	GECCO	}	&	Genetic and Evolutionary Computation Conference Companion	&	A	&	ERA	&	1	\\
\href{	https://2019.icse-conferences.org/track/icse-2019-Software-Engineering-in-Practice	}{	ICSE-SEIP	}	&	ICSE Software Engineering In Practice	&	A1	&	Qualis	&	1	\\
\href{	https://ieeeaccess.ieee.org/	}{	IEEE Access	}	&	IEEE Access	&	N/A	&	N/A	&	1	\\
\href{	https://dl.acm.org/journal/imwut	}{	IMWUT	}	&	Proceedings of the ACM on Interactive, Mobile, Wearable and Ubiquitous Technologies	&	N/A	&	N/A	&	1	\\
\href{	http://fcii.ute.edu.ec/inciscos/index.php/en	}{	INCISCOS	}	&	International Conference on Information Systems and Computer Science	&	N/A	&	N/A	&	1	\\
\href{	http://iv.csites.fct.unl.pt/	}{	IV	}	&	International Conference on Information Visualisation	&	B	&	ERA	&	1	\\
\href{	https://dl.acm.org/doi/proceedings/10.1145/2342509	}{	MCC	}	&	MCC workshop on Mobile cloud computing	&	N/A	&	N/A	&	1	\\
\href{	http://www.modelsward.org/	}{	MODELSWARD	}	&	International Conference on Model-Driven Engineering and Software Development	&	N/A	&	N/A	&	1	\\
\href{	http://www.msrconf.org/	}{	MSR	}	&	IEEE International Working Conference on Mining Software Repositories	&	A	&	ERA	&	1	\\
\href{	https://resources.sei.cmu.edu/library/asset-view.cfm?assetid=516150	}{	MTD	}	&	Workshop on Managing Technical Debt	&	N/A	&	N/A	&	1	\\
\href{	https://qrs19.techconf.org/	}{	QRS-C	}	&	IEEE International Conference on Software Quality, Reliability and Security Companion	&	B	&	ERA	&	1	\\
\href{	http://www.roots-conference.org/	}{	ROOTS	}	&	Reversing and Offensive-oriented Trends Symposium	&	N/A	&	N/A	&	1	\\
\href{	https://dblp.org/db/conf/sbes/index	}{	SBES	}	&	Brazilian Symposium on Software Engineering	&	B2	&	Qualis	&	1	\\
\href{	http://www.ieee-scam.org/	}{	SCAM	}	&	IEEE International Workshop on Source Code Analysis and Manipulation	&	C	&	ERA	&	1	\\
\href{	http://sccg.sk/	}{	SCCG	}	&	Spring Conference on Computer Graphics	&	C	&	ERA	&	1	\\
\href{	https://ieeexplore.ieee.org/xpl/RecentIssue.jsp?punumber=71	}{	TPDS	}	&	IEEE Transactions on Parallel and Distributed Systems	&	N/A	&	N/A	&	1	\\
\href{	https://conferences.computer.org/VLHCC/	}{	VL/HCC	}	&	IEEE Symposium on Visual Language and Human-Centric Computing	&	B	&	ERA	&	1	\\
\href{	https://dl.acm.org/conference/web3d	}{	Web3D	}	&	International ACM Conference on 3D Web Graphics and Interactive Technology	&	B1	&	Qualis	&	1	\\
\href{	https://www.wipsce.org/	}{	WiPSCE	}	&	Workshop in Primary and Secondary Computing Education	&	N/A	&	N/A	&	1	\\
  \bottomrule
  \end{tabu}%
\end{table}

We now elaborate on the results of software visualizations that we found in the research literature. We group the results based on the dimensions used in our classification. %\del{We found that 52 papers have been published in the IEEE International Workshop on Visualizing Software for Understanding and Analysis (VISSOFT). Other 36 papers have been published across 31 venues (as shown in Figure~\ref{fig:papers_venues})}\lm{this repeats what we wrote in "selected studies"}. Table~\ref{tab:papers_years} presents an overview of the publication years of the papers included in our study.
In the following, we present the results of our classifications using the six dimensions that we review (\ie software aspects, software processes, software engineering roles, information visualization techniques, mediums, and tools). %Here are the explanations in the texts. 
We present the results separated into two groups. Firstly, we present software engineering aspects. Those are, software aspects, software processes, and people in software engineering, which are presented in Tables~\ref{tab:aspects},~\ref{tab:software_process},and~\ref{tab:roles}. Secondly, we present information visualization aspects, which correspond to: information visualization techniques, display mediums, and tools. These aspects are presented in Tables~\ref{tab:infovis_techniques},~\ref{tab:mediums}, and~\ref{tab:tools}.
We also present exemplary software visualization tools (shown in Figures~\ref{fig:2013-2014},~\ref{fig:2015-2016}, and~\ref{fig:2017-2018}). We select examples that exhibit the highest number of citations.

\begin{table}
\parbox[t]{0.5\textwidth}{
\centering
  \begin{tabu}{p{0.075\textwidth}rrr}
  \toprule
   \textbf{Year} & \textbf{VISSOFT} & \textbf{Other Venues} & \textbf{Total}\\
  \midrule
  2013
& 7 & 9 & 16 \\
  2014
& 10 & 3 & 13 \\
  2015
& 10 & 8 & 18 \\
  2016 
& 7 & 5 & 12 \\
  2017
& 8 & 3 & 11 \\
  2018
& 10 & 8 & 18 \\
  2019
& 5 & 12 & 17 \\
\hline
  SUM
& 57 & 48 & 105 \\
  \bottomrule
  \end{tabu}%
\caption{Number of software visualization papers by year and venue.}%\lm{1) add a last row with the sum of vissoft papers for all years, for papers of other venues, and total, and 2) please create chars \mehf{what is chars?} for each of these tables (except table 6 in which the labels would be too long). Maybe those would work better.}}

\label{tab:papers_years}
}
\hfill
\parbox[t]{0.5\textwidth}{
\centering
  \begin{tabular}{p{0.25\textwidth}r}
  \toprule
   \textbf{Software Aspects}  & \textbf{Total} \\
  \midrule
Structure 
 & 34 \\
Behaviour 
 & 34 \\
Evolution 
 & 37 \\
  \bottomrule
  \end{tabular}%
\caption{Software aspects involved in software visualizations.}
\label{tab:aspects}
}
\vfill
\parbox[t]{0.5\textwidth}{
\centering
\begin{tabular}{p{0.25\textwidth}r}
  \toprule
   \textbf{Software Processes} &\textbf{Total} \\
  \midrule
Software design and implementation 
& 51 \\
Software maintenance  
& 24 \\
All software processes 
& 23 \\
Software validation 
& 7 \\
Software requirements 
& 0 \\
  \bottomrule
  \end{tabular}%
\caption{Software processes supported by software visualizations.}
\label{tab:software_process}
}
\hfill
\parbox[t]{0.5\textwidth}{
\centering
\begin{tabular}{p{0.25\textwidth}r}
  \toprule
   \textbf{Roles} & \textbf{Total}\\
  \midrule
Developer 
& 63 \\
Practitioner 
& 11 \\
Software Engineer
& 8 \\
Maintainer
& 5 \\
End-user
& 4 \\
Project Manager, Researcher, Tester
& 3 \\
Analysts, Architect
& 2 \\
Team Member
& 1 \\
Not Identified 
& 23 \\
  \bottomrule
  \end{tabular}%
\caption{Frequency of targeted roles played by users of software visualizations.}
 \label{tab:roles}
}
\vfill
\newcommand\tab[1][0.5cm]{\hspace*{#1}}
\parbox[t]{0.5\textwidth}{
\centering
\begin{tabu}{p{0.25\textwidth}r}
  \toprule
   \textbf{Information visualization techniques} & \textbf{Total}\\
  \midrule
  \emph{Empirical methodologies} &  \\
  \tab{Model} 
& 52\\
  \tab{Evaluation} 
 & 9\\
  \emph{Interactions} &  \\
  \tab{WIMP interactions} 
 & 5 \\
  \tab{Post-WIMP interactions} 
 & 5
 \\
    \emph{Frameworks} &  \\
  \tab{Systems and frameworks} 
 & 3\\
  \emph{Applications} &  \\
  \tab{Multivariate data visualization} 
& 13
 \\
  \tab{Graph visualization} 
 & 7
 \\
  \tab{Text visualization} 
 & 6
 \\
  \tab{Map visualization} 
 & 5
 \\
  \bottomrule
  \end{tabu}%
\caption{
Number of software visualizations grouped by information visualization techniques.}
\label{tab:infovis_techniques}
}
\hfill
\parbox[t]{0.5\textwidth}{
\centering
  \begin{tabu}{p{0.25\textwidth}r}
  \toprule
   \textbf{Mediums} & \textbf{Total}\\
  \midrule
  Standard Screen 
& 87 \\
  Immersive 3D Environment
& 8 \\
  Wall Display
& 2 \\
  Multi-Touch Table 
& 1 \\
Not Identified 
& 7 \\
  \bottomrule
  \end{tabu}%
\caption{Mediums used to display software visualizations.}
\label{tab:mediums}
}
\end{table}

\subsection{Software Engineering Aspects}
Software development is directly related to the software process. In the software process, there are many people involved in the process who are stakeholders of the software project.
Each phase in software development has multiple related software components. There are many aspects of software that project stakeholders need to understand, for which software visualization can be useful. Consequently, we discuss
\begin{inparaenum}[(\itshape i\upshape)]
\item  aspects of software systems, \item involved software process, and \item roles of stakeholders involved in software engineering. 
\end{inparaenum}
%\mehf{ come into the group of software visualization-wise.}

\subsubsection{Classification by software system aspects}
We classify the software visualizations based on the categories proposed by Diehl~\citep{diehl2007software}. His taxonomy classifies software visualizations based on the aspects of software that are supported. In it, software visualizations belong to one of three categories: 
\begin{inparaenum}[(\itshape i\upshape)]
  \item \emph{structure}, which includes visualizations that support the analysis of the static aspects and relationships in software systems, 
  \item \emph{behavior}, which relates to visualizations proposed for the analysis of data collected from the execution of programs, and
  \item \emph{evolution}, which contains visualizations that support to analyze how systems change over time.
\end{inparaenum}
Therefore, we analyze the number of papers that relate to each of these categories in the studied period.

Table~\ref{tab:aspects} shows that the number of papers that we found across the three aspects is almost balanced: 37 papers focus on the visualization of software evolution, 34 papers discuss the visualization of software behavior, and 34 papers describe visualizations of software structure.

% \begin{figure*}[htbp]
%  \centering 
%  \includegraphics[width=0.6\linewidth]{pictures/software_aspects.png}
%  \caption{Software aspects involved in software visualizations.}
%   \label{fig:aspects}
% \end{figure*}

The structure of software systems can be represented using a city metaphor (\eg buildings in the city represent classes in the software system). An example of a software visualization that uses the city metaphor is found in \emph{VR City}~\citep{vincur2017vr} (see Figure~\ref{fig:VRCity.png}). VR City enables users to observe and interact with a visual representation of source code that is displayed in a virtual reality environment. Another study~\citep{isaacs2018preserving} employs a directed acyclic graph to analyze dependencies of package management systems. \emph{Ravel}~\citep{isaacs2014combing} allows users to visualize the behaviour of a software system during the execution of a program based on log traces. In it, time is used to understand the parallelism in the execution of a program. Ravel allows users to search the event history to unveil insights. Users interact with the visualization using a standard computer screen, and mouse and keyboard. An example of a visualization that focus on code change history is \emph{Azurite}~\citep{yoon2013visualization}. In it, users can visualize the evolution of a software system that is displayed in a standard computer screen. Azurite is integrated into Eclipse as a plug-in. It enables developers and maintainers to go through the history of code via several editor commands.

\subsubsection{Classification by software engineering processes}
We classify selected papers by software processes based on a set of four process activities~\citep{sommerville2011software} that are fundamental to software engineering:
\begin{inparaenum}[(\itshape i\upshape)]
\item \emph{software specification} (specifically software requirements), which deals with the specification of software functionality and its constraints definition process.
\item \emph{software design and implementation}, which translates a specification into a concrete software product.
\item \emph{software validation} or \emph{software testing}, which prevents that a software system exhibits incorrect behavior.
\item \emph{software evolution} (specifically software maintenance), which deals with changes to a software system to deal with new requirements and discovered bugs that need to be fixed.
\end{inparaenum}
We organize the papers based on keywords that describe the name of a process, the abstract set of activities, and the tools introduced in those processes.

Table~\ref{tab:software_process} presents the results of our classification of papers by software process, we found: 51 papers that we classified into the design and implementation process, 24 papers that deal with the maintenance process, 23 papers that concern to all software processes, and 7 papers that focus on the software validation process. We \emph{did not} find visualizations proposed to support software requirements, though we are aware of the existence of such papers. Still, we consider that our results can give an overview of the degree of importance of each category.

% \begin{figure*}[htbp]
%  \centering 
%  \includegraphics[width=0.6\linewidth]{pictures/software_process.png}
%  \caption{Software processes supported by software visualizations.}
%   \label{fig:software_process}
% \end{figure*}

\begin{figure}[tbp]
\centering
\begin{subfigure}[t]{.5\textwidth}
  \centering
 \includegraphics[width=0.99\linewidth]{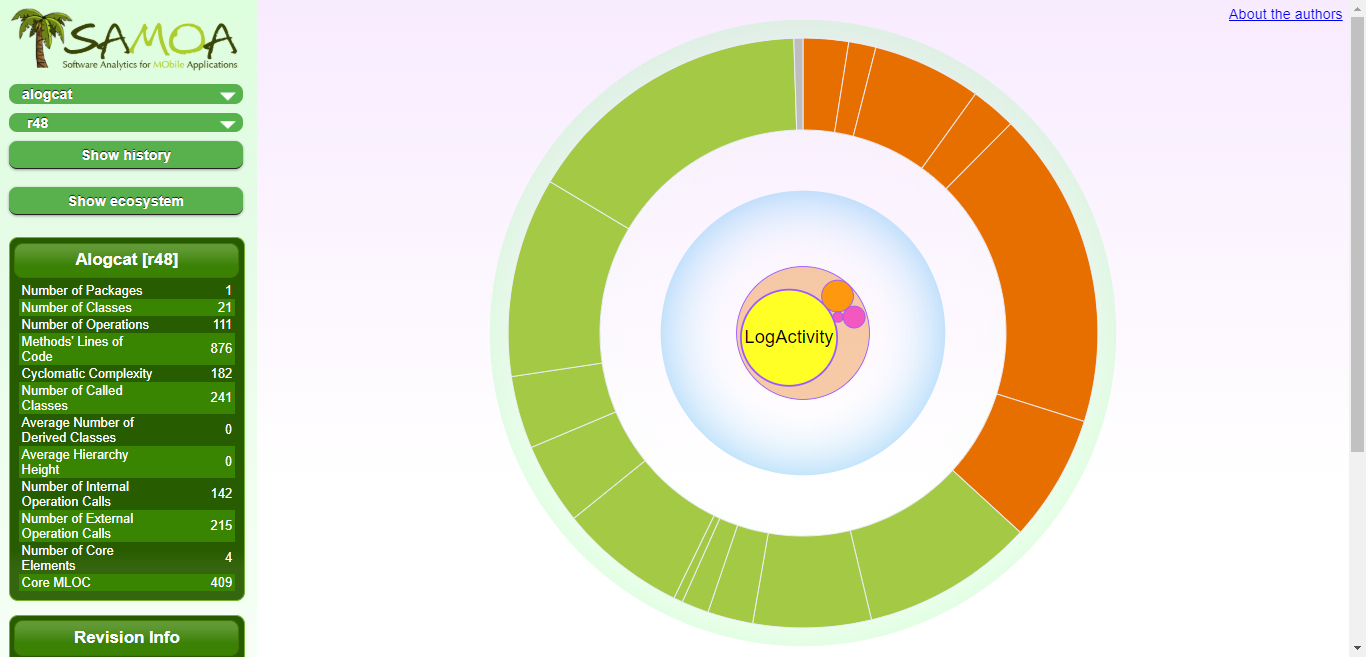}
 \caption{SAMOA, 2013}
  \label{fig:SAMOA.png}
\end{subfigure}%
\begin{subfigure}[t]{.5\textwidth}
  \centering
 \includegraphics[width=0.99\linewidth]{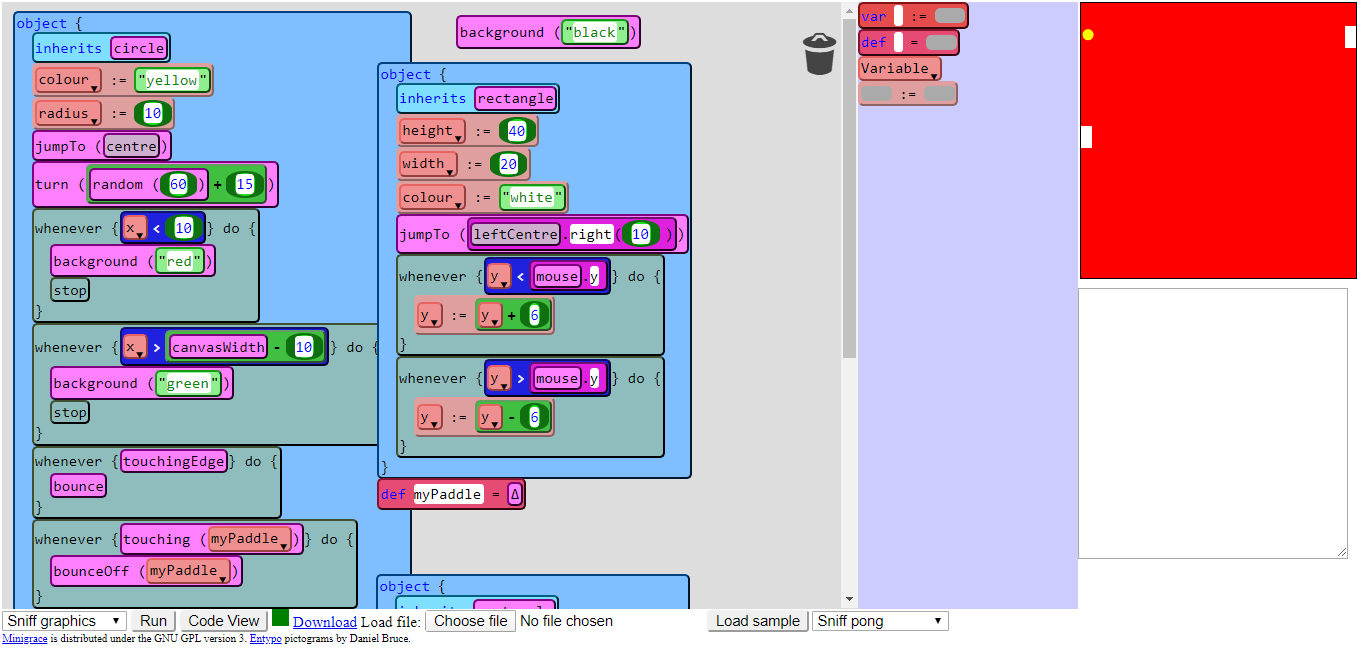}
 \caption{Tiled Grace, 2014}
 \label{fig:TiledGrace.PNG}
\end{subfigure}
\caption{\ref{fig:SAMOA.png}; \emph{SAMOA} displays the source code of the \emph{Alogcat} application. Figure taken from the Web~\citep{samoa2013web}, and reused with permission \textsuperscript{\textcopyright} 2013 Minelli.~\ref{fig:TiledGrace.PNG}; \emph{Tiled Grace} editing a small program in the ``sniff graphics'' dialect. Figure taken from the Web~\citep{tiledgrace2014web}, and reused with permission \textsuperscript{\textcopyright} 2014 Homer.}
%\caption{\ref{fig:SAMOA.png}; \emph{SAMOA} displays the source code of Alogcat application~\citep{minelli2013software}. \ref{fig:TiledGrace.PNG}; \emph{Tiled Grace}  animates transition from tiled to textual view from left to right~\citep{homer2014combining}.}
\label{fig:2013-2014}
\end{figure}

One study~\citep{benomar2013visualizing} uses a heatmap visualization to represent the execution and evolution of software during implementation. The task of this heatmap tool is to explore software dynamicity based on time and other dimensions of a software system. The tool is displayed on a standard computer screen.
Another study~\citep{gouveia2013using} propose three dynamic graphical forms, namely Sunburst, Vertical Partition, and Bubble Hierarchy to deal with fault detection in the verification process. The tool employs HTML5 to visualize software fault localization. The tool is offered as a plug-in for Eclipse and it is displayed on the standard computer screen.
\emph{CTRAS}~\citep{hao2019ctras} proposes a crowdsource testing method that takes advantage of code clones to enrich the content of bug descriptions and improve the efficiency of inspecting test reports through a comprehensive and comprehensible report.
\emph{SAMOA}~\citep{minelli2013software} (see Figure~\ref{fig:SAMOA.png}) is a visualization tool that supports software maintenance tasks. The tool visualizes source code, third-party libraries, and historical data using pie charts. The tool supports in-depth analysis of structural and evolutionary aspects of systems. The tools is implemented as a web application that is displayed on the standard computer screen.
One study~\citep{fronza2013cooperation} uses a Wordle visualization to support all software processes. The visualization tool, which is displayed on the computer screen, supports tasks that deal with the cooperation level of a development team using a Wordle-like visualization technique.

\subsubsection{Classification by software engineering roles}
We extracted the roles that envisioned users of software visualizations play in software engineering from frequent keywords in papers. In Table~\ref{tab:roles} we present the results of our classification of papers by software engineering roles. We found that proposed visualization envisioned ten different software engineering roles for their targeted users: Developer, Practitioner, Software Engineer, Maintainer, End-user, Project Manager, Researcher, Tester, Analysts, Architect, and Team Member. Notice that the second most frequent category corresponded to papers in which we did not identify an explicit role.

\begin{comment}
Moreover, we found the statistical data concerning people in software engineering. We separate into four groups such as one paper one role, one paper two roles, one paper more than two roles, and not identified.

The first group of one paper one role (57.95\%, \ie 51 papers), it seems to cover all of the software engineering processes from implementation to user. One paper one role who cooperate with software visualization and tool is developer, maintainer, software engineer, practitioner, architect, end-user, and test engineer respectively. The second group is one paper two roles (13.64\%, \ie 12 papers), it seems can play the umbrella activities, which based on two parts (management and engineering) of an \emph{ISO/IEC 29110} standard~\citep{ISO/IEC29110} that is engineering and management activities, in the software process, for example, the developer and project manager can use a visualization together. By the way, the developer and practitioner have the most mentioned (\ie 3 papers). Rest of them have one paper. Unlikely the first and the third group, this group has the researcher role work with another role. The third group is one paper more than one role (2.27\%, \ie 2 papers); the rest of them have three roles come together. Moreover, the last group is the not identified group has 26.14\% (\ie 23 papers).
\end{comment}

Roles such as developer, maintainer, and software engineer are directly related to the software implementation process and software maintenance process (and evidently related to all software processes), which are the top five processes mostly visualized. Interestingly, we found that the practitioner is the second largest number after the developer. Students, apprentices and new workers organize the practitioner.

% \begin{figure*}[htbp]
%  \centering 
%  \includegraphics[width=0.6\linewidth]{pictures/software_roles.png}
%  \caption{Frequency of targeted roles played by users of software visualizations.}
%   \label{fig:roles}
% \end{figure*}

One study~\citep{minelli2014visualizing} proposed a visualization, which is displayed on a standard computer screen, to analyze how developers use a graphical user interface of an integrated development environment. In it, developers need to understand and characterize development sessions with a timeline.
\emph{Variability Blueprint}~\citep{urli2015visual}, displayed on a computer screen, supports software product line (SPL) engineering through a visualization of feature models tree. The visualization tool helps software maintainers to understand dependencies between feature models.
\emph{RepoGrams}~\citep{rozenberg2016comparing} is a visualization tool for the analysis of software repositories (see Figure~\ref{fig:RepoGrams.PNG}). The tool, which is displayed on the standard computer screen, uses an extensible, metrics-based, visualization model as a footprint of the repository. The tool supports software engineers and researchers in comparative analyses of software projects over time.
\emph{Performance Evolution Matrix}~\citep{alcocer2019performance} is an interactive visualization that uses a matrix layout for the analysis of runtime metrics and source code changes, \eg execution graphs. The tool supports the analysis of performance metrics at different granularity levels of multiple versions of a software system. The study elaborates on evidence from an experiment that showed benefits of such interactive visualization for practitioners.

\subsection{Information Visualization}
We characterize aspects of software visualizations based on an employed information visualization technique, mediums, and tools.

\subsubsection{Classification by information visualization techniques}
Table~\ref{tab:infovis_techniques} presents our classification of the selected papers by information visualization technique based on four categories introduced in a previous study~\citep{liu2014survey}: 
\begin{inparaenum}[(\itshape i\upshape)]
\item \emph{empirical methodologies} usually correspond to novel visualization models and usability studies, \item \emph{interactions}  papers~\citep{yi2007toward} provide a comprehensive survey to study the role of interaction techniques in information visualization and the study, which can be further split into two categories: WIMP and post-WIMP interactions, \item \emph{systems and frameworks}, that is, \emph{systems} refer to toolkits for visualization construction, and \emph{frameworks} represent modeling of visualization techniques, and \item \emph{applications}, in which visualization designs are split into four groups based on the characteristics of the target data.
\end{inparaenum}

We found 61 software visualization papers that we classified as \emph{empirical methodologies}, 31 papers as \emph{applications}, 10 papers into the \emph{interactions} category, and 3 papers as \emph{systems and frameworks}.

\paragraph{Empirical methodologies}
%\subsubsection{Empirical methodologies}
Information visualization researchers have developed multiple empirical methods to support the design and implementation of novel visualization techniques, which can be classified in two subcategories: model and evaluation~\citep{liu2014survey}. Models are the basis of empirical research. Indeed, various models have been developed to assist in the design of effective data visualization. On the other hand, there are multiple challenges when evaluating software visualizations~\citep{merino2018systematic}. User studies are the most common method used in evaluations of information visualizations, which often involved measuring visualization performance. 
%We classify each paper by the main featured topic of the paper. For instance, most of the \emph{empirical methodologies} papers can be two kinds of its, which are model and evaluation, but for each paper has the structure like this, we will group it into the model at first. So it is the reason there are many papers in the model group. For the evaluation group, it includes the only paper that presents about evaluation methods and the special tools to evaluate some topics.

We found that amongst the 61 empirical methodologies papers, 52 papers correspond to the model category and 9 to evaluations. We observe there is an overlap between model and evaluation methods. %However, especially evaluation papers mentioned about only evaluation methods.

{CuboidMatrix}~\citep{schneider2016cuboidmatrix} is a visualization that uses a space-time metaphor to support users on understanding software evolution. Thus, users can navigate the space-time visualization to solve software comprehension tasks. The visualization is presented on a standard screen.

\begin{figure}[tbp]
\centering
\begin{subfigure}[t]{.5\textwidth}
  \centering
 \includegraphics[width=0.99\linewidth]{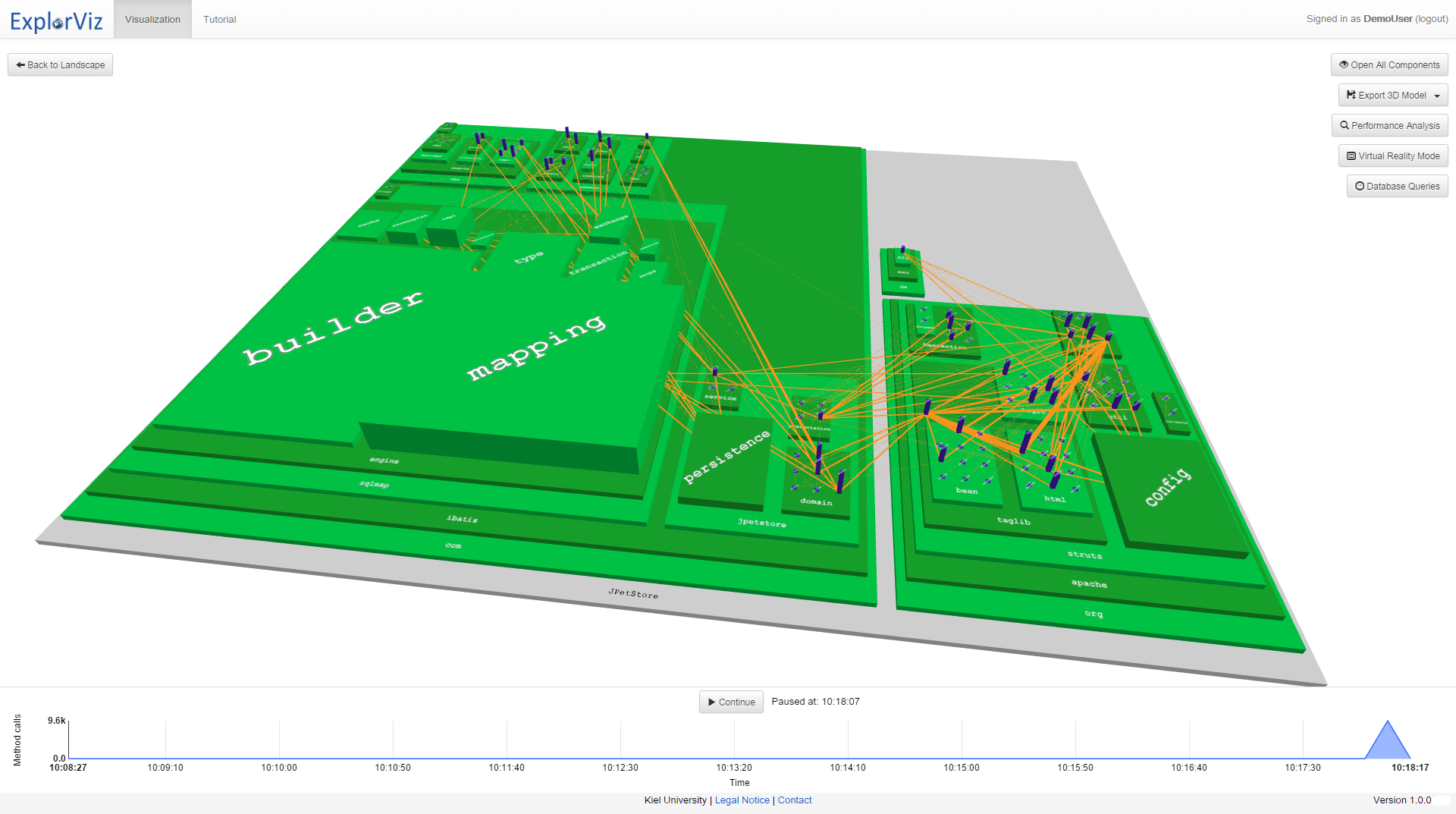}
 \caption{ExplorViz, 2015}
  \label{fig:ExplorViz.png}
\end{subfigure}%
\begin{subfigure}[t]{.5\textwidth}
  \centering
 \includegraphics[width=0.99\linewidth]{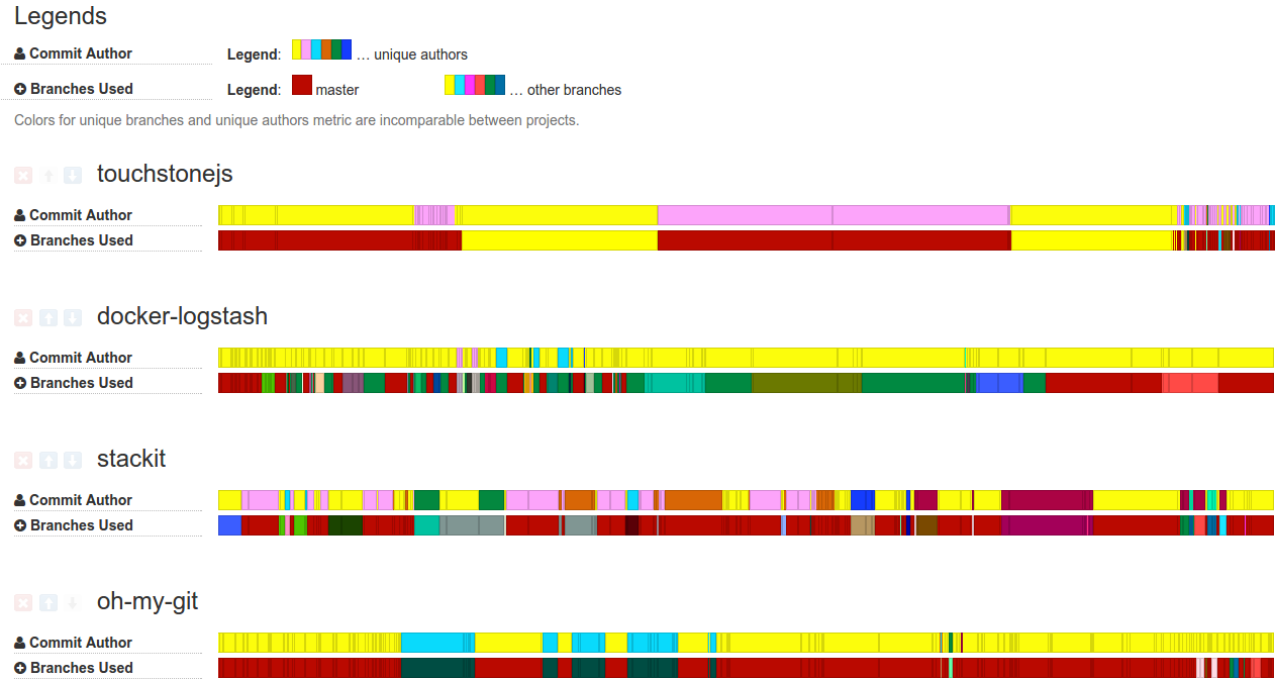}
 \caption{RepoGrams, 2016}
 \label{fig:RepoGrams.PNG}
\end{subfigure}
\caption{\ref{fig:ExplorViz.png}; \emph{ExploViz} supports the analysis of the architecture of a software system using a flat landscape 3D application visualization. Figure taken from the Web~\citep{explorviz2015web}, and reused with permission \textsuperscript{\textcopyright} 2015 Hasselbring.~\ref{fig:RepoGrams.PNG}; the \emph{RepoGrams} visualization shows eight software repository footprints, reused with permission \textsuperscript{\textcopyright} 2016 Rozenberg~\citep{rozenberg2016comparing}.
}
%\caption{\ref{fig:ExplorViz.png}; \emph{ExploViz} supports the analysis of the architecture of a software system using a flat landscape visualization~\citep{fittkau2015hierarchical}. \ref{fig:RepoGrams.PNG}; The \emph{RepoGrams} visualization shows eight software repository footprints~\citep{rozenberg2016comparing}.}
\label{fig:2015-2016}
\end{figure}

\emph{ExploViz}~\citep{fittkau2015hierarchical} is an example of visualization evaluation (see Figure~\ref{fig:ExplorViz.png}). In it, a model of the architecture of a software system is visualized using a flat landscape view. In the view, (1) green boxes represent nodes in the architecture and white labels are used to display the hostname of nodes; (2) purple boxes represent applications running amongst nodes; and (3) orange lines represent messages amongst applications. The visualization is presented on a standard screen. In a subsequent investigation, authors implemented a similar visualization displayed in virtual reality using a immersive 3D environment and 3D physical printed models.

\paragraph{Interactions}
%\subsubsection{Interactions}
User interaction is essential for data representation and data analysis. Interaction technologies can be classified in two categories: \emph{WIMP} interaction (\ie window, icon, mouse, pointer) and \emph{Post-WIMP} interactions, which is beyond traditional WIMP interaction, \eg reality-based interaction and 3D interaction. In papers that mention various categories, we identify a main interaction technique based on the content of the paper.

Amongst the interactions reported on papers, we found that 5 correspond to WIMP and 5 to post-WIMP interactions.  We observe that interactions often depend on the medium used to display visualizations. %Thanks for gadgets are emerging,  interactions (4.55\%, four papers) seem growing equal to WIMP interactions.

\emph{ConceptCloud}~\citep{greene2015interactive} combines a tag cloud visualization technique with a concept lattice to support source code navigation based on a flexible and interactive browser for \texttt{SVN} and \texttt{Git} repositories. This tool is an example of WIMP interactions. Visualizations are rendered on the standard screen.

\begin{figure}[tbp]
\centering
\begin{subfigure}[t]{.5\textwidth}
  \centering
 \includegraphics[width=0.99\linewidth]{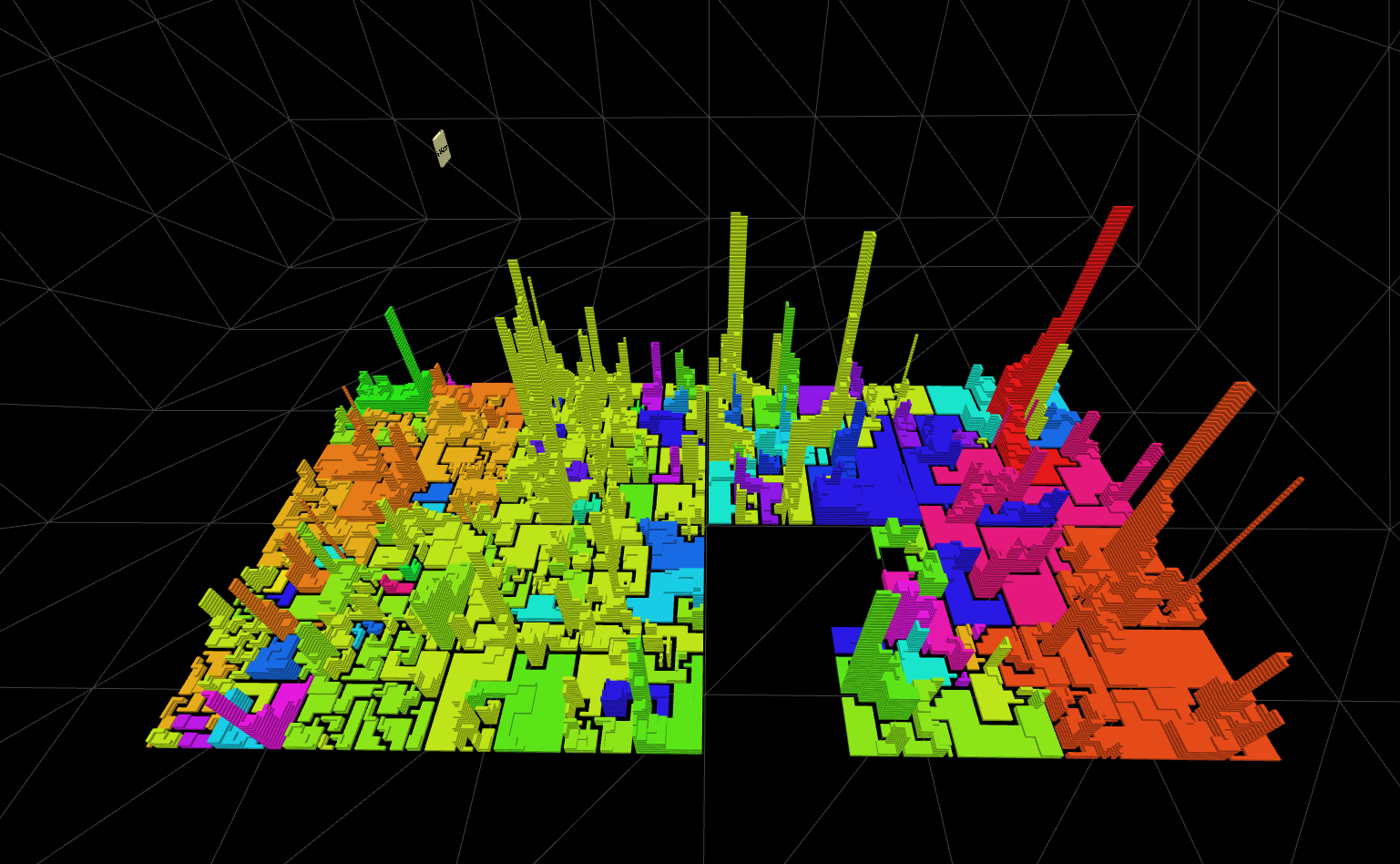}
 \caption{VR City, 2017}
  \label{fig:VRCity.png}
\end{subfigure}%
\begin{subfigure}[t]{.5\textwidth}
  \centering
 \includegraphics[width=0.99\linewidth]{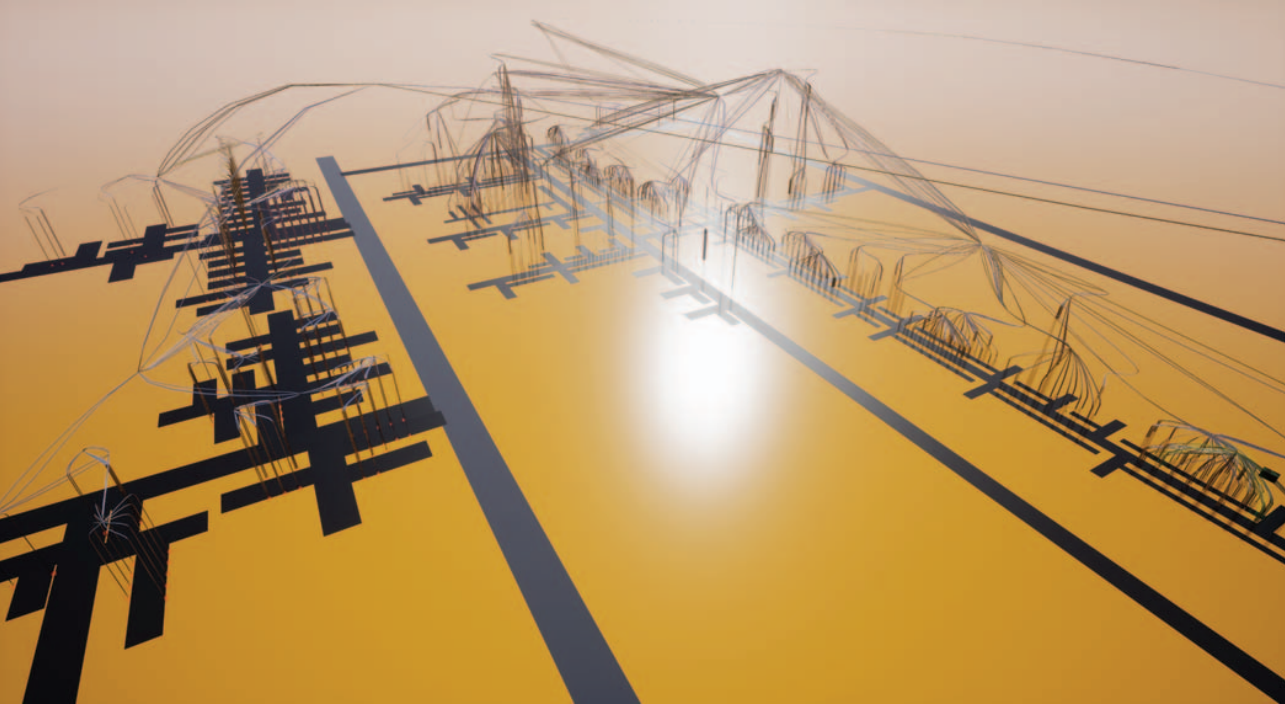}
 \caption{EvoStreets, 2018}
 \label{fig:EvoStreets.PNG}
\end{subfigure}
\caption{\ref{fig:VRCity.png}; The side view of \emph{VR City}, a city metaphor visualizes a software system in Virtual Reality environment, reused with permission \textsuperscript{\textcopyright} 2017 Vincur~\citep{vincur2017vr}.
\ref{fig:EvoStreets.PNG}; The bird’s eye view of \emph{EvoStreets} visualization, reprinted with permission \textsuperscript{\textcopyright} 2018 Koschke~\citep{rudel2018controlled}.
}
%\caption{\ref{fig:VRCity.png}; Views of the \emph{VR City} visualization for comparison of JHotDraw (top) and JUnit (bottom)~\citep{vincur2017vr}. \ref{fig:EvoStreets.PNG}; The bird’s eye view of \emph{EvoStreets} visualization~\citep{rudel2018controlled}.\lm{I would remove the citations in captions.}}
\label{fig:2017-2018}
\end{figure}

We also found a few post-WIMP interactions that often are used for visualizing software structure. \emph{VR City}~\citep{vincur2017vr}  employs virtual reality to display a software visualization that uses a city metaphor. Figure~\ref{fig:VRCity.png} presents views of the tool in which users can compare the characteristics of two software cities: JHotDraw (top) and JUnit (bottom). We observe that tjhe JUnit project looks smaller than JHotDraw, even though, it contains many more classes. Authors provide videos to obtain further details\footnote{\href{https://goo.gl/inrcZs}{https://goo.gl/inrcZs}}.

\paragraph{Frameworks}
%\subsubsection{Frameworks}
Systems and frameworks has only one subcategory. \emph{Systems} refer to libraries or toolkits for developing visualizations, whereas \emph{frameworks} are used to model various aspects of a visualization technique.  We found in this category only 2.86\% of 105 analyzed papers (\ie 3 papers. D3.js is probably the most popular library used to create Web interactive visualizations. Indeed, D3 is a powerful visualization framework  that can be used to create interactive visualizations based on HTML5, SVG, and CSS.

\emph{CodeCompass}~\citep{porkolab2018codecompass} is an open source LLVM/Clang-based tool developed by Ericsson Ltd. and Eötvös Loránd University, Budapest. The tool uses a variation of component diagrams which are implemented as graphs. The tool aims to help users understanding large legacy software systems based on static analysis and software metrics. \emph{CodeCompass} is a web-based plugin that is highly extensible.

\paragraph{Applications}
%\subsubsection{Applications}
To analyze the various types of data sets that can be involved in visualization, a study~\citep{liu2014survey} proposes 4 categories based on the characteristics of a data set. 
We found that nearly half of the papers in the applications category (\ie 31 papers) visualize multivariate data sets (\ie 13 papers). The rest can be split into graphs (\ie 7 papers), texts (\ie 6 papers), and maps (\ie 5 papers) respectively.

The \emph{applications} category take into account a visualization design that can involve an underlying graph, textual, map, and multivariate data. Therefore, we classified the list of the specific applications based on the underlying data and the visualization techniques. For example, one study~\citep{barik2014developers} elaborates on how developers comprehend compiler messages, which is a type of textual data, by providing a notification construction. Indeed, evaluation results are often used by developers to find explanations to development concerns. We group this work into the classification of text applications and behavior as the notation visualization is part of the textual data application.

Firstly, \emph{graph-like data} is related to topological structures such as dependencies amongst objects in an object-oriented systems or relationships among a group of people. \emph{Linvis}~\citep{wilde2018merge} is a visualization tool for the analysis of software repositories, which uses a conversion of directed acyclic graphs (DAG) to merged trees. This tool supports maintainers on the analysis of small projects. These methods and visualizations are implemented for the Web and displayed on the standard screen. 

Secondly, we observe that the visualization of \emph{textual data} aims at supporting the analysis of the semantics of data. For example, a visualization such that could helps software engineers to understand important concerns in an extensive collection of software project documents. An example of text visualization is the tile-based visualization named \emph{Tiled Grace}~\citep{homer2014combining} (see Figure~\ref{fig:TiledGrace.PNG}). Using drag-and-drop, developers interact with tiles to edit their programs, which they can complement with traditional textual environments. This tool is an educational programming language with a conventional textual syntax. This visualization is displayed on the standard screen.

Thirdly, visualizations to understand \emph{geological data} that relates space and size restricted to a terrestrial area. {CodeSurveyor}~\citep{hawes2015codesurveyor} is a visualization tool that uses a cartographic metaphor for creating a map of a code base. The interactive map allows users to zoom out and obtain a high-level overview of software components in which source files are shown as states in  continents. The visualization is displayed on the standard screen.  

Lastly, \emph{multivariate data} is a generic type that relates to a variety of fields. Usually, the goal of visualization of multivariate data is to explore the relationship amongst their dimensions. In many cases, the use of varying visualization techniques to understand inter-relationships is conducted by researchers and engineers. \emph{TraceDiff}~\citep{trumper2013multiscale} is a novel visualization method, which is based on icicle plots and hierarchical edge bundles. The approach uses a multiscale visualization metaphor to support the analysis of event traces, code structure, and function calls. The visualization is displayed on a standard screen.

\subsubsection{Classification by mediums}
Table~\ref{tab:mediums} presents the medium used to render software visualizations. A previous study~\citep{merino2018towards} analyzed the research literature on software visualization to extract development tasks, visualization techniques, and in particular, the mediums, which we adopted in our study as well. We confirmed that the standard screen is the most frequent medium used to display software visualization (\ie 87 papers). We also observe that in some papers authors do not explicitly mention the medium, though we infer it is a standard screen. Often, we infer the medium based on the analysis of figures in papers. Amongst the analyzed papers, we found a few approaches that used a medium other than the standard computer screen to display visualizations. In particular, we found 3D environments (\ie 8 papers), wall displays (\ie 2 papers), and multi-touch tables (\ie 1 paper) respectively. We also counted the number of papers in which we did not identify an explicit medium (\ie 7 papers).

\emph{SArF Map}~\citep{kobayashi2013sarf} visualizes components and layers of software systems using a city metaphor combined with a generated map. Developers and non-developers can use this visualization for high-level discussions and to make decisions in software development. This visualization  is displayed on a standard computer screen. \emph{EvoStreets}~\citep{rudel2018controlled} presents a software city metaphor visualization as well. Figure~\ref{fig:EvoStreets.PNG} shows a view of the visualization. Buildings in the visualization represent methods in the system, edges represent method calls, and streets represent package nesting and classes. This visualization is displayed in a 3D immersive virtual reality environment. \emph{Chronotwigger}~\citep{ens2014chronotwigger} enables the visualization of source and test files as visual nodes that dynamically change over time. This multi-user collaborative software visualization tool supports understanding source code and test code co-evolution. In it, users can specify a time span and select a node and zoom-in/out to analyze co-change. This visualization is displayed on a wall display combined with an immersive 3D environment. \emph{SourceVis}~\citep{anslow2013sourcevis} is a visualization tool designed to support multiple users, multiple visualization techniques, and to be displayed on a large shared interactive surface such as a multi-touch table. The aim of this tool is to understand how a system is structured based on the visualization of software metrics and source code evolution. This tool uses multi-touch tables for multi-user collaborative applications.

\subsubsection{Classification by tools}
Amongst the 105 reviewed papers, we found 62 papers in which we identified tools' names (see Table \ref{tab:tools}). That is, in only 28 papers we found an available link. When analyzing the evolution of the number of tools in the period 2013 -- 2019, we observe that the number of tools per year fluctuate between 3 (2013, 2014, 2017, 2019) and 6 (2016). We marked papers that contain available links with check mark (\checkmark) (and with dash (--) papers in which we did not found an available link). Moreover, tools are linked to URLs where they are available. Each tool is classified by:
\begin{inparaenum}[(\itshape i\upshape)] 
  \item Tool Names,
  \item References,
  \item Citation count\footnote{Last visit, Jan. 27, 2020},
  \item Software aspects (SoftVis),
  \item Software engineering process (\ie Software requirements [RE], Software Design and implementation [DI], Software validation [VV], Software maintenance [MA], All software processes [ALL]),
  \item Software Engineering Roles (Analyst [ANL], Architect [Arch], Developer [Dev], Maintainer [MTN], Practitioner [Prac], Project Manager [PM], Researcher {Res}, Software Engineer [SE], Team Member [Team], Tester [Test], End-user [User], Not Identified [N/A]), 
  \item Information visualization aspects (InfoVis)
  \item Medium (3D environments [3D], Multi-touch table/screen [MT], Standard screen [SS], Wall display [WL]; Not Identified [N/A]).
  \item Available code,
\end{inparaenum}
We think our list of available software visualization tools can offer a suitable complement to previous studies~\citep{merino2019vison}, and represent a means for practitioners who are willing to adopt software visualizations.

%\newcolumntype{R}{>{\raggedright\arraybackslash}m{7cm}}
\begin{table}[tbp]%[!ht]
  \caption{The list of software visualization tools organized by publication year and citation count. A asterisk mark is used to identify a publication year that differs between Google Scholar and IEEE Xplore.}
  \label{tab:tools}
  \setlength\tabcolsep{2pt}
 % \scriptsize%
  \centering%
  \begin{tabu}{llclllllc}
  \toprule
   \textbf{Tool Names} &\textbf{References} &\textbf{Citation} &\textbf{SoftVis} &\textbf{Process} &\textbf{Roles} &\textbf{InfoVis} &\textbf{Medium} &\textbf{Available}\\
  \midrule
\href{https://github.com/iris-42/CTRAS}{CTRAS}	&	\citep{hao2019ctras}	&	3	&	E	&	VV	&	Dev, Test	&	Model	&	SS	&	\checkmark	\\
{Performance evolution matrix}	&	\citep{alcocer2019performance}	&	1	&	E	&	MA	&	Dev, Prac	&	Model	&	SS	&	--	\\
{CloneCompass}	&	\citep{wang2019clonecompass}	&	0	&	E	&	MA	&	SE	&	Model	&	SS	&	--	\\
{CorpusVis}	&	\citep{slater2019corpusvis}	&	0	&	S	&	MA	&	Dev	&	Model	&	SS	&	--	\\
{Internal usage map}	&	\citep{anquetil2019decomposing}	&	0	&	S	&	MA	&	Dev	&	Model	&	SS	&	--	\\
{Evo-Clocks}	&	\citep{alexandru2019evo}	&	0	&	E	&	MA	&	Dev	&	Model	&	SS	&	--	\\
{EvoStreets (extended result)}	&	\citep{marcel2019movement}	&	0	&	S	&	ALL	&	Dev, Prac	&	Evaluation	&	3D	&	--	\\
{Atria}	&	\citep{williams2019visualizing}	&	0	&	B	&	DI	&	Dev	&	Model	&	SS	&	--	\\
\href{https://github.com/LLNL/Callflow}{Callflow}	&	\citep{nguyen2019visualizing}	&	0	&	B	&	ALL	&	Dev, ANL	&	Model	&	SS	&	\checkmark	\\
{City on the river (CotR)}	&	\citep{perrie2019city}	&	0	&	E	&	ALL	&	Dev, Team	&	Model	&	SS	&	--	\\
\href{https://github.com/santiontanon/Parallel}{Parallel}	&	\citep{zhu2019programming}	&	0	&	B	&	DI	&	Prac	&	Model	&	SS	&	\checkmark	\\
\href{https://github.com/Cloudslab/sdcon}{Clouds-Pi}	&	\citep{toosi2018clouds}	&	8	&	S	&	VV	&	N/A	&	Graph	&	N/A	&	\checkmark	\\
{EvoStreets}	&	\citep{rudel2018controlled}	&	6	&	S	&	ALL	&	Dev, Prac	&	Evaluation	&	3D	&	--	\\
{Feature visualiser}	&	\citep{duhoux2018feature}	&	6	&	B	&	DI	&	Dev	&	Model	&	N/A	&	--	\\
\href{https://github.com/Ericsson/CodeCompass}{CodeCompass}	&	\citep{porkolab2018codecompass}	&	3	&	E	&	ALL	&	Dev	&	Framework	&	SS	&	\checkmark	\\
\href{https://www.microsoft.com/en-us/research/project/cloudbuild-graph-explorer/}{BuildExplorer}	&	\citep{lebeuf2018understanding}	&	2	&	E	&	DI	&	Dev, SE	&	Model	&	SS	&	\checkmark	\\
{RepoVis}	&	\citep{feiner2018repovis}	&	2	&	E	&	DI	&	Dev, PM	&	Multivariate	&	SS	&	--	\\
{Parceive}	&	\citep{wilhelm2018tool}	&	1	&	B	&	DI	&	Dev	&	Evaluation	&	SS	&	--	\\
\href{https://doi.org/10.5281/zenodo.1311600}{Quality models inside out}	&	\citep{ulan2018quality}	&	1	&	E	&	MA	&	N/A	&	Multivariate	&	SS	&	\checkmark	\\
\href{https://remotion.cs.brown.edu/}{Remotion}	&	\citep{qian2018remotion}	&	0	&	E	&	ALL	&	ANL, User	&	Model	&	3D	&	\checkmark	\\
{VR city}	&	\citep{vincur2017vr}	&	24	&	S	&	DI	&	N/A	&	PostWIMP	&	3D	&	--	\\
{Code park}	&	\citep{khaloo2017code}	&	12	&	S	&	DI	&	Dev	&	WIMP	&	3D	&	--	\\
{CodeCity}	&	\citep{ogami2017using}	&	9	&	B	&	DI	&	Dev	&	Map	&	SS	&	--	\\
\href{https://fabian-beck.github.io/Method-Execution-Reports/}{Method execution reports}	&	\citep{beck2017method}	&	8	&	B	&	DI	&	Dev	&	Text	&	SS	&	\checkmark	\\
{TraceCompare}	&	\citep{doray2017diagnosing}	&	7	&	B	&	DI	&	Dev	&	Multivariate	&	N/A	&	--	\\
\href{https://github.com/SERESLab/iTrace-Archive/tree/issue88-vis}{iTraceVis}	&	\citep{clark2017itracevis}	&	6	&	S	&	DI	&	Dev, Res	&	Model	&	SS	&	\checkmark	\\
\href{https://sites.google.com/site/stefanhanenberg/software}{RegExVisualizer}	&	\citep{hollmann2017empirical}	&	3	&	S	&	DI	&	Prac	&	Evaluation	&	SS	&	\checkmark	\\
\href{http://repograms.net/}{RepoGrams}	&	\citep{rozenberg2016comparing}	&	17	&	E	&	MA	&	SE, Res	&	Multivariate	&	SS	&	\checkmark	\\
\href{http://dpf.hib.no}{WebDPF}	&	\citep{rabbi2016webdpf}	&	15	&	S	&	DI	&	N/A	&	Model	&	SS	&	\checkmark	\\
{Jsvee \& Kelmu}	&	\citep{sirkia2018jsvee}*	&	12	&	B	&	DI	&	Dev, Prac	&	Text	&	SS	&	--	\\
\href{https://github.com/mdfeist/TypeV}{TypeV}	&	\citep{feist2016visualizing}	&	9	&	E	&	ALL	&	Dev	&	Model	&	SS	&	\checkmark	\\
{CuboidMatrix}	&	\citep{schneider2016cuboidmatrix}	&	8	&	B	&	MA	&	Dev	&	Model	&	SS	&	--	\\
\href{http://dx.doi.org/10.5281/zenodo.56111}{Walls, pillars and beams}	&	\citep{tymchuk2016walls}	&	4	&	E	&	VV	&	Dev	&	Model	&	3D	&	\checkmark	\\
\href{http://li.turingmachine.org}{Linvis}	&	\citep{wilde2018merge}*	&	2	&	E	&	MA	&	MTN	&	Graph	&	SS	&	\checkmark	\\
\href{https://github.com/DeveloperLiberationFront/Excel-Function-Visualizer}{Perquimans}	&	\citep{middleton2016perquimans}	&	0	&	S	&	ALL	&	Prac, Res	&	Graph	&	SS	&	\checkmark	\\
\href{http://www.explorviz.net}{ExplorViz}	&	\citep{fittkau2015hierarchical}	&	19	&	S	&	DI	&	N/A	&	Evaluation	&	SS	&	\checkmark	\\
\href{http://www.conceptcloud.org/}{ConceptCloud}	&	\citep{greene2015interactive}	&	12	&	E	&	MA	&	Dev	&	WIMP	&	SS	&	\checkmark	\\
{CodeSurveyor}	&	\citep{hawes2015codesurveyor}	&	11	&	S	&	DI	&	Dev	&	Map	&	SS	&	--	\\
{Developer rivers}	&	\citep{burch2015visualizing}	&	11	&	E	&	ALL	&	Dev	&	Model	&	SS	&	--	\\
{Variability blueprint}	&	\citep{urli2015visual}	&	10	&	E	&	DI	&	MTN	&	Model	&	SS	&	--	\\
{Vimetrik}	&	\citep{khan2015visual}	&	5	&	E	&	MA	&	N/A	&	Model	&	SS	&	--	\\
{Goldenberry-GA}	&	\citep{garzon2015deconstructing}	&	3	&	S	&	MA	&	N/A	&	Model	&	SS	&	--	\\
\href{https://bitbucket.org/physviz/physviz}{PhysViz}	&	\citep{scarle2015visualising}	&	3	&	S	&	ALL	&	Dev	&	WIMP	&	SS	&	\checkmark	\\
\href{http://spideruci.github.io/cerebro}{Cerebro}	&	\citep{palepu2015revealing}	&	2	&	B	&	DI	&	SE	&	Model	&	SS	&	\checkmark	\\
\href{https://bitbucket.org/arthur486/jetracer}{JETracer}	&	\citep{molnar2015jetracer}	&	1	&	B	&	DI	&	Dev	&	Framework	&	SS	&	\checkmark	\\
{OctMiner}	&	\citep{lessa2015concern}	&	0	&	S	&	DI	&	N/A	&	Evaluation	&	SS	&	--	\\
\href{http://ecs.vuw.ac.nz/∼mwh/minigrace/tiled/}{Tiled grace}	&	\citep{homer2014combining}	&	45	&	B	&	DI	&	Dev	&	Text	&	SS	&	\checkmark	\\
{Feature relations graphs (FRoGs)}	&	\citep{martinez2014feature}	&	32	&	E	&	MA	&	N/A	&	Model	&	SS	&	--	\\
{AniMatrix}	&	\citep{rufiange2014animatrix}	&	26	&	E	&	MA	&	Arch	&	WIMP	&	SS	&	--	\\
{ChronoTwigger}	&	\citep{ens2014chronotwigger}	&	17	&	E	&	ALL	&	Dev, Test	&	PostWIMP	&	WL	&	--	\\
\href{https://github.com/kuleszdl/SIFEI}{SIFEI}	&	\citep{kulesz2014integrating}	&	9	&	E	&	ALL	&	N/A	&	Model	&	SS	&	\checkmark	\\
{AR map}	&	\citep{dugerdil2014visualizing}	&	5	&	S	&	DI	&	MTN	&	Model	&	SS	&	--	\\
\href{http://www.sts.tu-harburg.de/projects/regvis/regvis.html}{regVIS}	&	\citep{toprak2014lightweight}	&	4	&	S	&	DI	&	N/A	&	Model	&	SS	&	\checkmark	\\
\href{http://samoa.inf.usi.ch/}{Samoa}	&	\citep{minelli2013software}	&	108	&	S	&	MA	&	N/A	&	Multivariate	&	SS	&	\checkmark	\\
\href{http://www.cs.cmu.edu/~azurite/}{Azurite}	&	\citep{yoon2013visualization}	&	57	&	E	&	MA	&	Dev	&	Model	&	SS	&	\checkmark	\\
{SourceVis}	&	\citep{anslow2013sourcevis}	&	36	&	E	&	DI	&	Dev	&	PostWIMP	&	MT	&	--	\\
{TraceDiff}	&	\citep{trumper2013multiscale}	&	31	&	B	&	DI	&	N/A	&	Multivariate	&	SS	&	--	\\
{Performance evolution blueprint}	&	\citep{alcocer2013performance}	&	31	&	E	&	MA	&	N/A	&	Multivariate	&	SS	&	--	\\
{SArF map}	&	\citep{kobayashi2013sarf}	&	29	&	S	&	DI	&	Dev, User	&	Map	&	SS	&	--	\\
{SyncTrace}	&	\citep{karran2013synctrace}	&	28	&	B	&	DI	&	N/A	&	Multivariate	&	SS	&	--	\\
\href{http://www.cosc.canterbury.ac.nz/research/RG/svg/taggle}{Taggle}	&	\citep{emerson2013toy}	&	9	&	S	&	DI	&	Dev	&	Model	&	SS	&	\checkmark	\\
{SoftDynamik}	&	\citep{grznar2013visualizing}	&	4	&	B	&	DI	&	Dev	&	Model	&	SS	&	--	\\
  \bottomrule
  \end{tabu}%
\end{table}

\section{Discussion}
\label{sec:discussion}
%aspects
In the following section, we discuss on the results of the topics described previously, \ie Software system aspects, Software engineering processes, Software engineering roles, Information visualization techniques, Mediums, and Tools.

\textbf{Software Systems Aspects.} We found that the number of software visualizations proposed to support each of the three aspects of systems (\ie structure, behavior, and evolution) is almost balanced. We found that 32.39\% (\ie 34 papers) of the analyzed papers describe visualizations proposed for the analysis of the structure of software systems. We found that more than a third of papers, \ie 35.24\%, (\ie 37 papers) focus on visualizations of the evolution. We also found that 32.39\% (\ie 34 papers) of the 105 reviewed papers relate to visualizations proposed to analyze the behavior of software systems. 

We think that visualizations proposed to deal with questions that relate to these aspects are a good fit to the increasing complexity of software systems. We expect our review can be used by practitioners to obtain an overview of modern software visualization tools, some of them available, that have been proposed in the research literature. However, if we look at the number of available tools, we observe that the lowest number of tools are present in the \emph{Behavior} (\ie 6 tools), then the \emph{Structure} aspect in which we found  10 available tools, and finally, the \emph{Evolution} aspect in which available tools are a bit more frequent (\ie 12 tools). %It can be seen that although the overall number of research works is almost equal, in the \emph{Behavior} aspect has a tiny amount of access. It may be challenging to extend the work in this aspect.

%processes
\textbf{Software Engineering Processes.} We found that 48.58\% of papers (\ie 51 papers) relate to the design and implementation process. That is, we found that 22.86\% of papers (\ie 24 papers) related to visualizations proposed to support tasks of the maintenance process. We found that 21.91\% of papers (\ie 23 papers) relate to all processes in the software life-cycle, and that 6.67\% of papers (\ie 7 papers) relate to software testing process. We did not find visualizations proposed to support software requirements process, possibly due our employed method to collect primary studies. Of course, it is possible that software visualization papers that relate to the software requirements process are published, for instance, as short papers. %In the future, we will expand the framework of making a systematic literature review wider. 
In summary, we found that proposed software visualizations mostly support tasks involved with the design and implementation processes. 
Software maintenance is another frequent process in which we classified many proposed visualizations. We observe that aspects of evolution significantly relate to the design and implementation and maintenance process. In general, we observe there are multiple visualizations proposed to support various phases of software life-cycle. We found only a few visualizations that support tasks which involve the software testing process. We ask why there are only a few software visualizations especially designed to support software requirements engineering and software testing processes. Although we understand there is an intrinsic complexity in these subjects, we think these processes can represent an opportunity for future research. 
One reason that explains the many visualizations proposed to support all processes is that software development requires communicating several processes which, in some cases, may impose a challenge. In particular, the challenge of define a clear separation between processes, for which, visualization can be suitable by showing, for instance, an overview and progress of development. In addition, software is inherently tangible. %In consequence, the design and implementation processes, which are fundamental necessary process to change the need to be useful software. Therefore, w
We think that visualization can play an important role, in particular, to support the design and implementation processes. Indeed, we observe that the number of visualization that supports these processes is significant. Amongst papers that present visualization to support these processes, we found 12 available tools for design and implementation process. We also found 6 available visualization tools for maintenance and 7 tools for all software engineering processes. We only found 3 available visualization tools that support software testing tasks. We observe that these findings are consistent with that ``developer'' is the most frequent targeted software engineering role.

%roles
\textbf{Software Engineering Roles.} The envisioned user of a software visualization tool plays a particular role as a stakeholder in the life-cycle of a software system. We analyzed roles described in the reviewed papers to understand how well defined are these envisioned roles. We found that \emph{developer} is the most frequently envisioned role of the users of software visualizations ($\times$12). Proposed software visualizations that target the developer audience, usually focus on supporting source code understanding. The developers role crosscuts the three aspects of software (\ie structure, behavior, evolution) that are visualized. Developers need to understand the structure (\eg to modify a legacy software system), the behavior (\eg to identify issues with the execution of programs), and the evolution (\eg to analyze how software projects evolve). However, we also observe that developers in real-world projects usually play more specific roles (\eg analysts, testers).
We believe that the high frequency of generic roles in the expected audience of software visualizations can be a symptom of a disconnect with the reality in real-world software projects~\citep{merino2018towards}. We suggest researchers in the software visualization field to identify the specific context of the targeted users of their proposed visualizations. We also recommend to identify specific users roles by analyzing real-world software projects.
Other less frequent roles ($\times$1--3) that play envisioned users of software visualizations are researchers, software engineers, users, practitioners, analysts and maintainers. Of course, developers can require multiple tools to understand a software system. %Still, compared to the total number of researches reaching the developer (\ie 50 times) and the number of available tools (\ie 12 times), it is considered a small amount. However, the other positions that need to understand the software as well. There are very few supporting tools. 
We consider this to be an opportunity for developing such tools, which could be used for education and industrial work.

%InfoVis
\textbf{Information Visualization Techniques.} We classified information visualization techniques that we found amongst the analyzed software visualization papers. In them, we found \emph{empirical methodologies} (58.10\%, \ie 61 papers), \emph{applications} (29.53\%, \ie 31 papers), \emph{interactions} (9.53\%, \ie 10 papers), and \emph{system and frameworks} (2.86\%, \ie 3 papers). we observe that \emph{empirical methodologies} is the most frequent technique. We think that the needs of information visualization researchers and software engineers of developing empirical methods to support the design and implementation of novel software visualizations can explain it. Another reason that explains it is that ecosystems in which software systems execute have multiple characteristics which are, usually, hard to understand. To support such analysis, visualization can be helpful. The least frequent category is ``system and framework'', which corresponds to models and artifacts used to develop applications. %We observe that developers use multiple frameworks to creating the visualization application. The aim of software visualization may need to focus on how to understand the software more than creating a novel framework to develop the software visualization application.

\begin{figure}[htbp]
 \centering 
 \includegraphics[width=\linewidth]{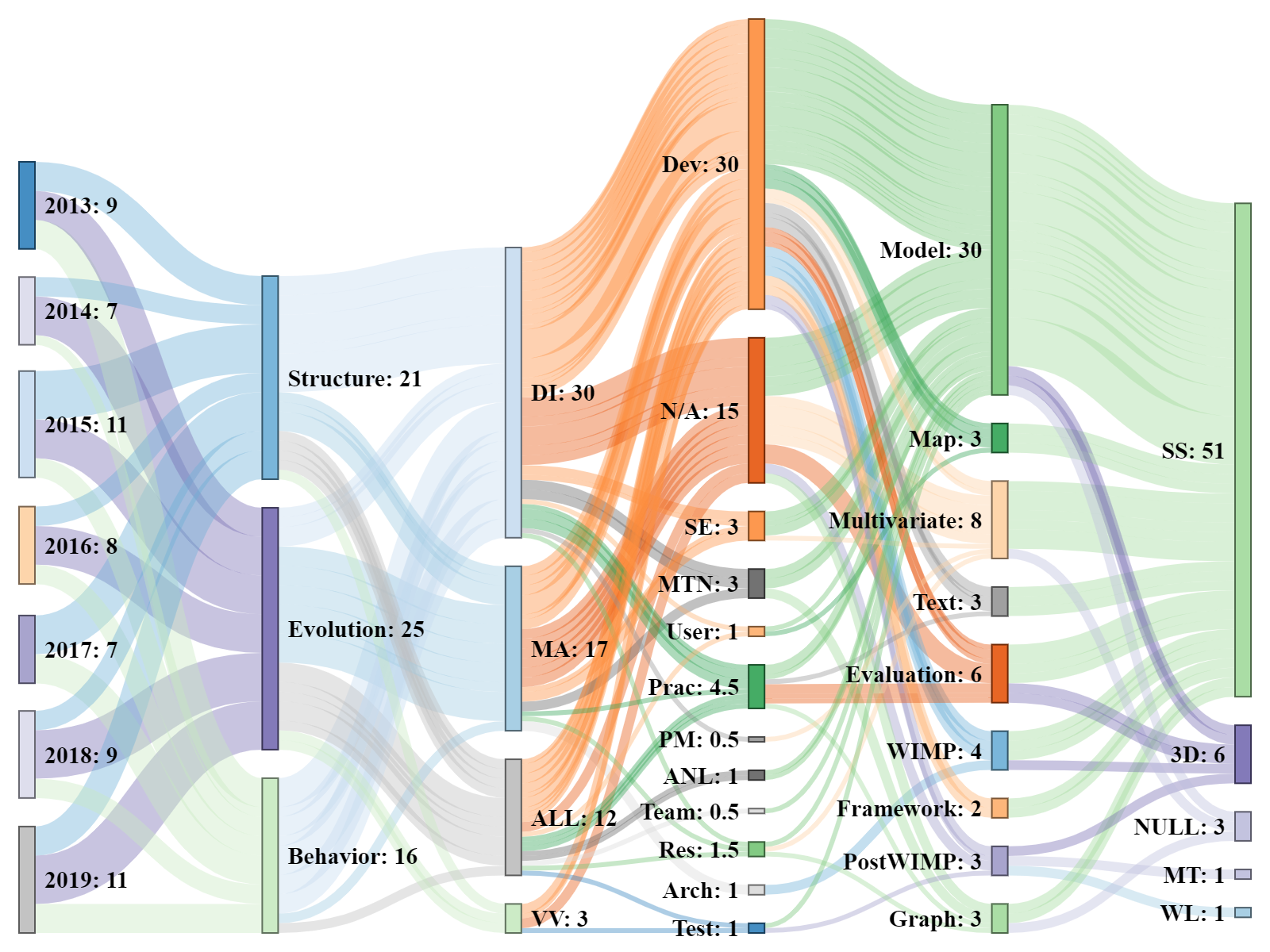}
 \caption{A Sankey diagram that shows the distribution and relationships between software visualization papers published in 2013--2019. The levels in the diagram represent: year, software aspects, software processes, roles, information visualization techniques, and mediums, respectively.}
  \label{fig:sankey2013_2019}
\end{figure}

%mediums
\textbf{Mediums.} The most frequent medium used to render software visualization is the standard computer screen (82.86\%, \ie 87 papers), 3D environments (7.62\%, \ie 8 papers), wall displays (1.91\%, \ie 2 papers), and multi-touch tables (0.96\%, \ie 1 papers). We found that even though the majority of software visualizations` are displayed on the standard computer screen, a few other media have been used as well. The analysis of the evolution of the various media used to display software visualizations shows a trend that suggests that in the future we could expect that many proposed visualizations are going to be displayed in a medium other than the standard computer screen. For instance, we envision that with the rising of agile development, in which a team needs to discuss their work or present results to a steering committee in a meeting or war room, visualizations displayed in modern media will be used more frequently. Though we observe that these other modern media (\eg virtual and augmented reality devices) require more support for users to interact and collaborate. For example, a visualization tool, such as  ExplorViz, has been continuously improved. Recently, a report~\citep{zirkelbach2019hands} describe hand gestures added to interact with the software visualization in VR. %\lm{do we have a figure or table that supports this analysis? If so, make a reference to it}

%tools
\textbf{Tools.} we found multiple software visualization tools mentioned amongst papers, however, we observe that almost half of them are not available. We also observe that the average lifespan of software visualization tools is about 3.7 years \citep{merino2016towards}. The reason for the short lifespan is one thing that should be considered for future research.

%Sankey
The Sankey diagram showed in Figure~\ref{fig:sankey2013_2019} presents the relationship amongst the characteristics of software visualizations, which are grouped into three aspects of software: Evolution, Structure, and Behavior.
We observe that most of the visualizations support structural and behavioral aspects, which are often related to the design and implementation process. Also, the developer role is frequently envisioned as the target user of software visualizations. We observe that such role relates to the Model category using InfoVis techniques to support the design and implementation processes. Furthermore, most the standard computer screen is often the medium used to display software visualizations. We also notice that design and implementation processes heavily relate to  the practitioner role who are associated to the evaluation in InfoVis the category. We conjecture that this could be explained as it is often necessary to measure and Evaluate the use of trainees in understanding software through software visualization tools.
Software visualizations that support the analysis of evolution often support tasks involved in the maintenance process, though often the envisioned user is described as developer as opposed to maintainer, which could be explained in a lack of clear definition of the scope of the maintainer role.
We observe that visualizations that support the software testing process focus on the evolution and structure aspects of systems. Similarly, the envisioned users of such visualizations are often developers and not testers. %It may be possible that the job description and job title are also unclear, like the developer and the maintainer above.}

In this paper, we report on the state-of-the-art of software visualization published between 2013--2019. We recommend to readers, possibly practitioners, interested in consulting software visualization published in previous years to access the software visualization blog \url{https://softvis.wordpress.com/papers/overviews/}.
 
%%%%%%%%%%%%%%%%%%%%%% CONCLUSION %%%%%%%%%%%%%%%%%%%%%%
\section{Conclusion}
\label{sec:conclusion}
In this paper, we provide an overview of the state-of-the-art of software visualization. We reviewed 105 software visualization papers published during the past six years. We concentrated on system papers as they often describe software visualization tools. System papers can be classified based on six aspects. From each paper, we extracted:
\begin{inparaenum}[(\itshape i\upshape)] 
  \item aspects of software visualization (\ie structure, behavior, evolution),
  \item supported software engineering processes,
  \item roles of expected users,
  \item media used to display visualizations,
  \item employed visualization techniques, and
  \item provide the software visualization tools list.
\end{inparaenum}

%aspects
We found that %Though all the state-of-the-art of software visualization supports ternary aspects of software, a bit over 
about a third of the analyzed papers support software evolution tasks.
%process and roles
We found the envisioned users of proposed software visualizations have multiple roles. We observe an increasing number of software visualizations proposed to support all software engineering processes, which could be due the fact that this process involves users that play multiple roles.
%medium
The majority of software visualizations are displayed using a standard computer screen; however, we do observe that several approaches that support the analysis of the structure of systems are displayed in 3D environments with city metaphor. Interestingly, we found that the rate of proposed software visualizations that are displayed in a medium different than the computer screen is increasing consistently. 
%vis technique
When analyzing employed information visualization techniques, we observe that empirical methodologies are frequently used to support software engineering concerns. In addition, we also found that whereas the number of empirical studies conducted to evaluate software visualizations has gradually increased, software visualizations that support the analysis of multivariate data has increased as well.
%tools
We found that the average lifespan of software visualization tools is 3.7 years. Amongst the identified software visualizations tools, we observe that only half of them are available. %One reason is the lack of software maintenance which is the main reason that is generally related to the lifetime of the software.

We think our investigation can help developers to identify software visualization tools for concrete development tasks, for instance, a software development team could use storylines~\citep{tang2018istoryline} to represent software development activity, as well as, a rich text visualization~\citep{cao2010facetatlas} for visualizing the software requirements.

In the future, we plan to expand our investigation by integrating the analysis of events and activities~\citep{wu2018streamexplorer} and social media activity~\citep{zhao2014fluxflow,cao2012whisper}. We also plan to investigate how visualizations displayed in immersive 3D environments such as head-mounted displays for virtual and augmented reality can help to increase the awareness of software requirements throughout the development life-cycle. We hope information visualization researchers, developers, and software engineering can use our classification to identify suitable visualization for their particular context in software projects.

\begin{acknowledgements}
This work is supported by National Natural Science Foundation of China (61772463, 61772456, 61761136020). Merino acknowledges funding by the Deutsche Forschungsgemeinschaft (DFG, German Research Foundation) -- Project-ID 251654672 -- TRR 161. We thank Wilhelm Hasselbring, Michael Homer, Rainer Koschke, Roberto Minelli, Daniel Rozenberg, and Juraj Vincur, and for permitting us to reuse exemplary figures of their software visualizations.
\end{acknowledgements}


%\bibliographystyle{chicago}
\bibliographystyle{spbasic}      % basic style, author-year citations
\bibliography{template}   % name your BibTeX data base

\begin{thebibliography}{106}
\providecommand{\natexlab}[1]{#1}
\providecommand{\url}[1]{{#1}}
\providecommand{\urlprefix}{URL }
\expandafter\ifx\csname urlstyle\endcsname\relax
  \providecommand{\doi}[1]{DOI~\discretionary{}{}{}#1}\else
  \providecommand{\doi}{DOI~\discretionary{}{}{}\begingroup
  \urlstyle{rm}\Url}\fi
\providecommand{\eprint}[2][]{\url{#2}}

\bibitem[{Alexandru et~al.(2019)Alexandru, Proksch, Behnamghader, and
  Gall}]{alexandru2019evo}
Alexandru CV, Proksch S, Behnamghader P, Gall HC (2019) {Evo-Clocks}: Software
  evolution at a glance. In: IEEE Working Conference on Software Visualization
  (VISSOFT), IEEE, pp 12--22

\bibitem[{Anquetil et~al.(2019)Anquetil, Etien, Andreo, and
  Ducasse}]{anquetil2019decomposing}
Anquetil N, Etien A, Andreo G, Ducasse S (2019) Decomposing god classes at
  siemens. In: IEEE International Conference on Software Maintenance and
  Evolution (ICSME), IEEE, pp 169--180

\bibitem[{Anslow et~al.(2013)Anslow, Marshall, Noble, and
  Biddle}]{anslow2013sourcevis}
Anslow C, Marshall S, Noble J, Biddle R (2013) {SourceVis}: Collaborative
  software visualization for co-located environments. In: IEEE Working
  Conference on Software Visualization (VISSOFT), IEEE, pp 1--10

\bibitem[{Barik et~al.(2014)Barik, Lubick, Christie, and
  Murphy-Hill}]{barik2014developers}
Barik T, Lubick K, Christie S, Murphy-Hill E (2014) How developers visualize
  compiler messages: A foundational approach to notification construction. In:
  IEEE Working Conference on Software Visualization (VISSOFT), IEEE, pp 87--96

\bibitem[{Beck et~al.(2017{\natexlab{a}})Beck, Burch, Diehl, and
  Weiskopf}]{beck2017taxonomy}
Beck F, Burch M, Diehl S, Weiskopf D (2017{\natexlab{a}}) A taxonomy and survey
  of dynamic graph visualization. Computer Graphics Forum 36(1):133--159

\bibitem[{Beck et~al.(2017{\natexlab{b}})Beck, Siddiqui, Bergel, and
  Weiskopf}]{beck2017method}
Beck F, Siddiqui HA, Bergel A, Weiskopf D (2017{\natexlab{b}}) Method execution
  reports: Generating text and visualization to describe program behavior. In:
  IEEE Working Conference on Software Visualization (VISSOFT), IEEE, pp 1--10

\bibitem[{Benomar et~al.(2013)Benomar, Sahraoui, and
  Poulin}]{benomar2013visualizing}
Benomar O, Sahraoui H, Poulin P (2013) Visualizing software dynamicities with
  heat maps. In: IEEE Working Conference on Software Visualization (VISSOFT),
  IEEE, pp 1--10

\bibitem[{Bergel and Beck(2017)}]{bergel2017guest}
Bergel A, Beck F (2017) Guest editorial of the special section on software
  visualization. Information and Software Technology 87:221--222

\bibitem[{Burch et~al.(2015)Burch, Munz, Beck, and
  Weiskopf}]{burch2015visualizing}
Burch M, Munz T, Beck F, Weiskopf D (2015) Visualizing work processes in
  software engineering with developer rivers. In: IEEE Working Conference on
  Software Visualization (VISSOFT), IEEE, pp 116--124

\bibitem[{Cao et~al.(2010)Cao, Sun, Lin, Gotz, Liu, and Qu}]{cao2010facetatlas}
Cao N, Sun J, Lin YR, Gotz D, Liu S, Qu H (2010) Facetatlas: Multifaceted
  visualization for rich text corpora. IEEE Transactions on Visualization and
  Computer Graphics 16(6):1172--1181

\bibitem[{Cao et~al.(2012)Cao, Lin, Sun, Lazer, Liu, and Qu}]{cao2012whisper}
Cao N, Lin YR, Sun X, Lazer D, Liu S, Qu H (2012) Whisper: Tracing the
  spatiotemporal process of information diffusion in real time. IEEE
  Transactions on Visualization and Computer Graphics 18(12):2649--2658

\bibitem[{Clark and Sharif(2017)}]{clark2017itracevis}
Clark B, Sharif B (2017) {iTraceVis}: Visualizing eye movement data within
  eclipse. In: IEEE Working Conference on Software Visualization (VISSOFT),
  IEEE, pp 22--32

\bibitem[{Diehl(2007)}]{diehl2007software}
Diehl S (2007) Software visualization: visualizing the structure, behaviour,
  and evolution of software. Springer Science \& Business Media

\bibitem[{Doray and Dagenais(2017)}]{doray2017diagnosing}
Doray F, Dagenais M (2017) Diagnosing performance variations by comparing
  multi-level execution traces. IEEE Transactions on Parallel and Distributed
  Systems 28(2):462--474

\bibitem[{Dugerdil and Niculescu(2014)}]{dugerdil2014visualizing}
Dugerdil P, Niculescu M (2014) Visualizing software structure
  understandability. In: Australian Software Engineering Conference (ASWEC),
  IEEE, pp 110--119

\bibitem[{Duhoux et~al.(2018)Duhoux, Mens, and Dumas}]{duhoux2018feature}
Duhoux B, Mens K, Dumas B (2018) Feature visualiser: an inspection tool for
  context-oriented programmers. In: International Workshop on Context-Oriented
  Programming: Advanced Modularity for Run-time Composition, ACM, pp 15--22

\bibitem[{Emerson et~al.(2013)Emerson, Churcher, and Deaker}]{emerson2013toy}
Emerson J, Churcher N, Deaker C (2013) From toy to tool: Extending tag clouds
  for software and information visualisation. In: Australian Software
  Engineering Conference (ASWEC), IEEE, pp 155--164

\bibitem[{Ens et~al.(2014)Ens, Rea, Shpaner, Hemmati, Young, and
  Irani}]{ens2014chronotwigger}
Ens B, Rea D, Shpaner R, Hemmati H, Young JE, Irani P (2014) {ChronoTwigger}: A
  visual analytics tool for understanding source and test co-evolution. In:
  IEEE Working Conference on Software Visualization (VISSOFT), IEEE, pp
  117--126

\bibitem[{Feiner and Andrews(2018)}]{feiner2018repovis}
Feiner J, Andrews K (2018) {RepoVis}: Visual overviews and full-text search in
  software repositories. In: IEEE Working Conference on Software Visualization
  (VISSOFT), IEEE, pp 1--11

\bibitem[{Feist et~al.(2016)Feist, Santos, Watts, and
  Hindle}]{feist2016visualizing}
Feist MD, Santos EA, Watts I, Hindle A (2016) Visualizing project evolution
  through abstract syntax tree analysis. In: IEEE Working Conference on
  Software Visualization (VISSOFT), IEEE, pp 11--20

\bibitem[{Fittkau et~al.(2015{\natexlab{a}})Fittkau, Krause, and
  Hasselbring}]{fittkau2015hierarchical}
Fittkau F, Krause A, Hasselbring W (2015{\natexlab{a}}) Hierarchical software
  landscape visualization for system comprehension: A controlled experiment.
  In: IEEE Working Conference on Software Visualization (VISSOFT), IEEE, pp
  36--45

\bibitem[{Fittkau et~al.(2015{\natexlab{b}})Fittkau, Zirkelbach, Krause, and
  Hasselbring}]{explorviz2015web}
Fittkau F, Zirkelbach C, Krause A, Hasselbring W (2015{\natexlab{b}})
  Explorviz. \urlprefix\url{https://www.explorviz.net/}, accessed February 25,
  2020

\bibitem[{Fronza et~al.(2013)Fronza, Janes, Sillitti, Succi, and
  Trebeschi}]{fronza2013cooperation}
Fronza I, Janes A, Sillitti A, Succi G, Trebeschi S (2013) Cooperation wordle
  using pre-attentive processing techniques. In: International Workshop on
  Cooperative and Human Aspects of Software Engineering (CHASE), IEEE, pp
  57--64

\bibitem[{Garz{\'o}n-Rodriguez et~al.(2015)Garz{\'o}n-Rodriguez, Diosa, and
  Rojas-Galeano}]{garzon2015deconstructing}
Garz{\'o}n-Rodriguez LP, Diosa HA, Rojas-Galeano S (2015) Deconstructing {GAs}
  into visual software components. In: Annual Conference on Genetic and
  Evolutionary Computation, ACM, pp 1125--1132

\bibitem[{Geisler(1998)}]{geisler1998making}
Geisler G (1998) Making information more accessible: A survey of information
  visualization applications and techniques. URL: http://www ils unc edu/\~{}
  geisg/info/infovis/paper html

\bibitem[{Gouveia et~al.(2013)Gouveia, Campos, and Abreu}]{gouveia2013using}
Gouveia C, Campos J, Abreu R (2013) Using {HTML5} visualizations in software
  fault localization. In: IEEE Working Conference on Software Visualization
  (VISSOFT), IEEE, pp 1--10

\bibitem[{Greene and Fischer(2015)}]{greene2015interactive}
Greene GJ, Fischer B (2015) Interactive tag cloud visualization of software
  version control repositories. In: IEEE Working Conference on Software
  Visualization (VISSOFT), IEEE, pp 56--65

\bibitem[{Greene et~al.(2017)Greene, Esterhuizen, and
  Fischer}]{greene2017visualizing}
Greene GJ, Esterhuizen M, Fischer B (2017) Visualizing and exploring software
  version control repositories using interactive tag clouds over formal concept
  lattices. Information and Software Technology 87:223--241

\bibitem[{Grzn{\'a}r and Kapec(2013)}]{grznar2013visualizing}
Grzn{\'a}r F, Kapec P (2013) Visualizing dynamics of object oriented programs
  with time context. In: Spring Conference on Computer Graphics, ACM, pp 65--72

\bibitem[{Hao et~al.(2019)Hao, Feng, Jones, Li, and Chen}]{hao2019ctras}
Hao R, Feng Y, Jones JA, Li Y, Chen Z (2019) {CTRAS}: Crowdsourced test report
  aggregation and summarization. In: IEEE/ACM International Conference on
  Software Engineering (ICSE), IEEE, pp 900--911

\bibitem[{Hawes et~al.(2015)Hawes, Marshall, and
  Anslow}]{hawes2015codesurveyor}
Hawes N, Marshall S, Anslow C (2015) {CodeSurveyor}: Mapping large-scale
  software to aid in code comprehension. In: IEEE Working Conference on
  Software Visualization (VISSOFT), IEEE, pp 96--105

\bibitem[{Hollmann and Hanenberg(2017)}]{hollmann2017empirical}
Hollmann N, Hanenberg S (2017) An empirical study on the readability of regular
  expressions: Textual versus graphical. In: IEEE Working Conference on
  Software Visualization (VISSOFT), IEEE, pp 74--84

\bibitem[{Homer and Noble(2014{\natexlab{a}})}]{homer2014combining}
Homer M, Noble J (2014{\natexlab{a}}) Combining tiled and textual views of
  code. In: IEEE Working Conference on Software Visualization (VISSOFT), IEEE,
  pp 1--10

\bibitem[{Homer and Noble(2014{\natexlab{b}})}]{tiledgrace2014web}
Homer M, Noble J (2014{\natexlab{b}}) Tiled grace.
  \urlprefix\url{https://homepages.ecs.vuw.ac.nz/~mwh/minigrace/tiled/},
  accessed February 25, 2020

\bibitem[{Isaacs and Gamblin(2018)}]{isaacs2018preserving}
Isaacs KE, Gamblin T (2018) Preserving command line workflow for a package
  management system using ascii dag visualization. IEEE Transactions on
  Visualization and Computer Graphics 25(9):2804--2820

\bibitem[{Isaacs et~al.(2014)Isaacs, Bremer, Jusufi, Gamblin, Bhatele, Schulz,
  and Hamann}]{isaacs2014combing}
Isaacs KE, Bremer PT, Jusufi I, Gamblin T, Bhatele A, Schulz M, Hamann B (2014)
  Combing the communication hairball: Visualizing parallel execution traces
  using logical time. IEEE Transactions on Visualization and Computer Graphics
  20(12):2349--2358

\bibitem[{Karran et~al.(2013)Karran, Trumper, and
  D\"{o}llner}]{karran2013synctrace}
Karran B, Trumper J, D\"{o}llner J (2013) Synctrace: Visual thread-interplay
  analysis. In: IEEE Working Conference on Software Visualization (VISSOFT),
  IEEE, pp 1--10

\bibitem[{Khaloo et~al.(2017)Khaloo, Maghoumi, Taranta, Bettner, and
  Laviola}]{khaloo2017code}
Khaloo P, Maghoumi M, Taranta E, Bettner D, Laviola J (2017) {Code Park}: A new
  {3D} code visualization tool. In: IEEE Working Conference on Software
  Visualization (VISSOFT), IEEE, pp 43--53

\bibitem[{Khan et~al.(2015)Khan, Barthel, Ebert, and
  Liggesmeyer}]{khan2015visual}
Khan T, Barthel H, Ebert A, Liggesmeyer P (2015) Visual analytics of software
  structure and metrics. In: IEEE Working Conference on Software Visualization
  (VISSOFT), IEEE, pp 16--25

\bibitem[{Kienle and Muller(2007)}]{kienle2007requirements}
Kienle HM, Muller HA (2007) Requirements of software visualization tools: A
  literature survey. In: IEEE International Workshop on Visualizing Software
  for Understanding and Analysis (VISSOFT), IEEE, pp 2--9

\bibitem[{Kitchenham et~al.(2002)Kitchenham, Pfleeger, Pickard, Jones, Hoaglin,
  Emam, and Rosenberg}]{kitchenham2000guidelines}
Kitchenham BA, Pfleeger SL, Pickard LM, Jones PW, Hoaglin DC, Emam KE,
  Rosenberg J (2002) Preliminary guidelines for empirical research in software
  engineering. IEEE Transactions on Software Engineering 22(8):721--734,
  \doi{10.1109/TSE.2002.1027796}

\bibitem[{Kobayashi et~al.(2013)Kobayashi, Kamimura, Yano, Kato, and
  Matsuo}]{kobayashi2013sarf}
Kobayashi K, Kamimura M, Yano K, Kato K, Matsuo A (2013) {SArF} map:
  Visualizing software architecture from feature and layer viewpoints. In: IEEE
  International Conference on Program Comprehension (ICPC), IEEE, pp 43--52

\bibitem[{Kulesz et~al.(2014)Kulesz, Scheurich, and
  Beck}]{kulesz2014integrating}
Kulesz D, Scheurich J, Beck F (2014) Integrating anomaly diagnosis techniques
  into spreadsheet environments. In: IEEE Working Conference on Software
  Visualization (VISSOFT), IEEE, pp 11--19

\bibitem[{Kumar(2016)}]{kumar2016review}
Kumar S (2016) A review of recent trends and issues in visualization.
  International Journal on Computer Science and Engineering (IJCSE) 8(3):41--54

\bibitem[{Lebeuf et~al.(2018)Lebeuf, Voyloshnikova, Herzig, and
  Storey}]{lebeuf2018understanding}
Lebeuf C, Voyloshnikova E, Herzig K, Storey MA (2018) Understanding, debugging,
  and optimizing distributed software builds: A design study. In: IEEE
  International Conference on Software Maintenance and Evolution (ICSME), IEEE,
  pp 496--507

\bibitem[{Lessa et~al.(2015)Lessa, Carneiro, Monteiro, and
  e~Abreu}]{lessa2015concern}
Lessa IDM, Carneiro GDF, Monteiro MP, e~Abreu FB (2015) A concern visualization
  approach for improving {MATLAB} and octave program comprehension. In:
  Brazilian Symposium on Software Engineering (SBES), IEEE, pp 130--139

\bibitem[{Liu et~al.(2014)Liu, Cui, Wu, and Liu}]{liu2014survey}
Liu S, Cui W, Wu Y, Liu M (2014) A survey on information visualization: recent
  advances and challenges. The Visual Computer 30(12):1373--1393

\bibitem[{Liu et~al.(2017)Liu, Wang, Liu, and Zhu}]{liu2017towards}
Liu S, Wang X, Liu M, Zhu J (2017) Towards better analysis of machine learning
  models: A visual analytics perspective. Visual Informatics 1(1):48--56

\bibitem[{Maletic et~al.(2002)Maletic, Marcus, and Collard}]{maletic2002task}
Maletic JI, Marcus A, Collard ML (2002) A task oriented view of software
  visualization. In: International Workshop on Visualizing Software for
  Understanding and Analysis (VISSOFT), IEEE, pp 32--40

\bibitem[{Martinez et~al.(2014)Martinez, Ziadi, Mazo, Bissyand{\'e}, Klein, and
  Le~Traon}]{martinez2014feature}
Martinez J, Ziadi T, Mazo R, Bissyand{\'e} TF, Klein J, Le~Traon Y (2014)
  Feature relations graphs: A visualisation paradigm for feature constraints in
  software product lines. In: IEEE Working Conference on Software Visualization
  (VISSOFT), IEEE, pp 50--59

\bibitem[{Mattila et~al.(2016)Mattila, Ihantola, Kilamo, Luoto, Nurminen, and
  V{\"a}{\"a}t{\"a}j{\"a}}]{mattila2016software}
Mattila AL, Ihantola P, Kilamo T, Luoto A, Nurminen M, V{\"a}{\"a}t{\"a}j{\"a}
  H (2016) Software visualization today: Systematic literature review. In:
  International Academic Mindtrek Conference, ACM, pp 262--271

\bibitem[{McNabb and Laramee(2017)}]{mcnabb2017survey}
McNabb L, Laramee RS (2017) Survey of surveys ({SoS})-mapping the landscape of
  survey papers in information visualization. Computer Graphics Forum
  36(3):589--617

\bibitem[{Merino et~al.(2016)Merino, Ghafari, and
  Nierstrasz}]{merino2016towards}
Merino L, Ghafari M, Nierstrasz O (2016) Towards actionable visualisation in
  software development. In: IEEE Working Conference on Software Visualization
  (VISSOFT), IEEE, pp 61--70

\bibitem[{Merino et~al.(2018{\natexlab{a}})Merino, Ghafari, Anslow, and
  Nierstrasz}]{merino2018systematic}
Merino L, Ghafari M, Anslow C, Nierstrasz O (2018{\natexlab{a}}) A systematic
  literature review of software visualization evaluation. Journal of Systems
  and Software 144:165--180

\bibitem[{Merino et~al.(2018{\natexlab{b}})Merino, Ghafari, and
  Nierstrasz}]{merino2018towards}
Merino L, Ghafari M, Nierstrasz O (2018{\natexlab{b}}) Towards actionable
  visualization for software developers. Journal of Software: Evolution and
  Process 30(2):e1923

\bibitem[{Merino et~al.(2019)Merino, Kozlova, Nierstrasz, and
  Weiskopf}]{merino2019vison}
Merino L, Kozlova E, Nierstrasz O, Weiskopf D (2019) {VISON}: An ontology-based
  approach for software visualization tool discoverability. In: IEEE Working
  Conference on Software Visualization (VISSOFT), IEEE, pp 45--55

\bibitem[{Middleton and Murphy-Hill(2016)}]{middleton2016perquimans}
Middleton J, Murphy-Hill E (2016) Perquimans: A tool for visualizing patterns
  of spreadsheet function combinations. In: IEEE Working Conference on Software
  Visualization (VISSOFT), IEEE, pp 51--60

\bibitem[{Minelli and Lanza(2013{\natexlab{a}})}]{samoa2013web}
Minelli R, Lanza M (2013{\natexlab{a}}) Samoa.
  \urlprefix\url{http://samoa.inf.usi.ch/}, accessed February 25, 2020

\bibitem[{Minelli and Lanza(2013{\natexlab{b}})}]{minelli2013software}
Minelli R, Lanza M (2013{\natexlab{b}}) Software analytics for mobile
  applications--insights and lessons learned. In: European Conference on
  Software Maintenance and Reengineering (CSMR), IEEE, pp 144--153

\bibitem[{Minelli et~al.(2014)Minelli, Mocci, Lanza, and
  Baracchi}]{minelli2014visualizing}
Minelli R, Mocci A, Lanza M, Baracchi L (2014) Visualizing developer
  interactions. In: IEEE Working Conference on Software Visualization
  (VISSOFT), IEEE, pp 147--156

\bibitem[{Molnar(2015)}]{molnar2015jetracer}
Molnar AJ (2015) {JETracer} a framework for {Java} {GUI} event tracing. In:
  International Conference on Evaluation of Novel Approaches to Software
  Engineering (ENASE), IEEE, pp 207--214

\bibitem[{Mumtaz et~al.(2019)Mumtaz, Latif, Beck, and
  Weiskopf}]{mumtaz2019exploranative}
Mumtaz H, Latif S, Beck F, Weiskopf D (2019) Exploranative code quality
  documents. IEEE transactions on visualization and computer graphics
  26(1):1129--1139

\bibitem[{Munzner(2008)}]{munzner2008process}
Munzner T (2008) Process and pitfalls in writing information visualization
  research papers. In: International Conference on Information Visualisation
  (IV), Springer, pp 134--153

\bibitem[{{Nguyen} et~al.(2019){Nguyen}, {Bhatele}, {Jain}, {Kesavan},
  {Bhatia}, {Gamblin}, {Ma}, and {Bremer}}]{nguyen2019visualizing}
{Nguyen} HTP, {Bhatele} A, {Jain} N, {Kesavan} S, {Bhatia} H, {Gamblin} T, {Ma}
  K, {Bremer} P (2019) Visualizing hierarchical performance profiles of
  parallel codes using callflow. IEEE Transactions on Visualization and
  Computer Graphics pp 1--1, \doi{10.1109/TVCG.2019.2953746}

\bibitem[{Ogami et~al.(2017)Ogami, Kula, Hata, Ishio, and
  Matsumoto}]{ogami2017using}
Ogami K, Kula RG, Hata H, Ishio T, Matsumoto K (2017) Using high-rising cities
  to visualize performance in real-time. In: IEEE Working Conference on
  Software Visualization (VISSOFT), IEEE, pp 33--42

\bibitem[{Palepu and Jones(2015)}]{palepu2015revealing}
Palepu VK, Jones JA (2015) Revealing runtime features and constituent behaviors
  within software. In: IEEE Working Conference on Software Visualization
  (VISSOFT), IEEE, pp 86--95

\bibitem[{Perrie et~al.(2019)Perrie, Xie, Nayebi, Fokaefs, Lyons, and
  Stroulia}]{perrie2019city}
Perrie J, Xie J, Nayebi M, Fokaefs M, Lyons K, Stroulia E (2019) City on the
  river: visualizing temporal collaboration. In: International Conference on
  Computer Science and Software Engineering, pp 82--91

\bibitem[{Porkol{\'a}b et~al.(2018)Porkol{\'a}b, Brunner, Krupp, and
  Csord{\'a}s}]{porkolab2018codecompass}
Porkol{\'a}b Z, Brunner T, Krupp D, Csord{\'a}s M (2018) Codecompass: an open
  software comprehension framework for industrial usage. In: International
  Conference on Program Comprehension (ICPC), ACM, pp 361--369

\bibitem[{Qian et~al.(2018)Qian, Chapin, Papoutsaki, Yang, Nelissen, and
  Huang}]{qian2018remotion}
Qian J, Chapin A, Papoutsaki A, Yang F, Nelissen K, Huang J (2018) Remotion: A
  motion-based capture and replay platform of mobile device interaction for
  remote usability testing. ACM on Interactive, Mobile, Wearable and Ubiquitous
  Technologies 2(2):77

\bibitem[{Rabbi et~al.(2016)Rabbi, Lamo, Yu, and Kristensen}]{rabbi2016webdpf}
Rabbi F, Lamo Y, Yu IC, Kristensen LM (2016) {WebDPF}: A web-based
  metamodelling and model transformation environment. In: International
  Conference on Model-Driven Engineering and Software Development (MODELSWARD),
  IEEE, pp 87--98

\bibitem[{Rodrigues-Jr et~al.(2015)Rodrigues-Jr, Zaina, Oliveira, Brandoli, and
  Traina}]{rodrigues2015survey}
Rodrigues-Jr J, Zaina L, Oliveira M, Brandoli B, Traina A (2015) A survey on
  information visualization in light of vision and cognitive sciences. arXiv
  preprint arXiv:150507079

\bibitem[{Rozenberg et~al.(2016)Rozenberg, Beschastnikh, Kosmale, Poser,
  Becker, Palyart, and Murphy}]{rozenberg2016comparing}
Rozenberg D, Beschastnikh I, Kosmale F, Poser V, Becker H, Palyart M, Murphy GC
  (2016) Comparing repositories visually with repograms. In: IEEE/ACM Working
  Conference on Mining Software Repositories (MSR), IEEE, pp 109--120

\bibitem[{R\"{u}del et~al.(2018)R\"{u}del, Ganser, and
  Koschke}]{rudel2018controlled}
R\"{u}del MO, Ganser J, Koschke R (2018) A controlled experiment on spatial
  orientation in {VR}-based software cities. In: IEEE Working Conference on
  Software Visualization (VISSOFT), IEEE, pp 21--31

\bibitem[{Rufiange and Melan{\c{c}}on(2014)}]{rufiange2014animatrix}
Rufiange S, Melan{\c{c}}on G (2014) Animatrix: A matrix-based visualization of
  software evolution. In: IEEE Working Conference on Software Visualization
  (VISSOFT), IEEE, pp 137--146

\bibitem[{Sandoval~Alcocer et~al.(2013)Sandoval~Alcocer, Bergel, Ducasse, and
  Denker}]{alcocer2013performance}
Sandoval~Alcocer JP, Bergel A, Ducasse S, Denker M (2013) {Performance
  Evolution Blueprint}: Understanding the impact of software evolution on
  performance. In: IEEE Working Conference on Software Visualization (VISSOFT),
  IEEE, pp 1--9

\bibitem[{Sandoval~Alcocer et~al.(2019)Sandoval~Alcocer, Beck, and
  Bergel}]{alcocer2019performance}
Sandoval~Alcocer JP, Beck F, Bergel A (2019) {Performance Evolution Matrix}:
  Visualizing performance variations along software versions. In: IEEE Working
  Conference on Software Visualization (VISSOFT), IEEE, pp 1--11

\bibitem[{Scarle and Walkinshaw(2015)}]{scarle2015visualising}
Scarle S, Walkinshaw N (2015) Visualising software as a particle system. In:
  IEEE Working Conference on Software Visualization (VISSOFT), IEEE, pp 66--75

\bibitem[{Schneider et~al.(2016)Schneider, Tymchuk, Salgado, and
  Bergel}]{schneider2016cuboidmatrix}
Schneider T, Tymchuk Y, Salgado R, Bergel A (2016) {CuboidMatrix}: Exploring
  dynamic structural connections in software components using space-time cube.
  In: IEEE Working Conference on Software Visualization (VISSOFT), IEEE, pp
  116--125

\bibitem[{Sensalire et~al.(2008)Sensalire, Ogao, and
  Telea}]{sensalire2008classifying}
Sensalire M, Ogao P, Telea A (2008) Classifying desirable features of software
  visualization tools for corrective maintenance. In: ACM Symposium on Software
  Visualization (SOFTVIS), ACM, pp 87--90

\bibitem[{Shahin et~al.(2014)Shahin, Liang, and Babar}]{shahin2014systematic}
Shahin M, Liang P, Babar MA (2014) A systematic review of software architecture
  visualization techniques. Journal of Systems and Software 94:161--185

\bibitem[{Sirki{\"a}(2018)}]{sirkia2018jsvee}
Sirki{\"a} T (2018) {Jsvee} \& {Kelmu}: Creating and tailoring program
  animations for computing education. Journal of Software: Evolution and
  Process 30(2):e1924

\bibitem[{Slater et~al.(2019)Slater, Anslow, Dietrich, and
  Merino}]{slater2019corpusvis}
Slater J, Anslow C, Dietrich J, Merino L (2019) {CorpusVis}--visualizing
  software metrics at scale. In: IEEE Working Conference on Software
  Visualization (VISSOFT), IEEE, pp 99--109

\bibitem[{Sommerville(2011)}]{sommerville2011software}
Sommerville I (2011) Software engineering. Person Education Ltd

\bibitem[{Steinbeck et~al.(2019)Steinbeck, Koschke, and
  R\"{u}del}]{marcel2019movement}
Steinbeck M, Koschke R, R\"{u}del MO (2019) Movement patterns and trajectories
  in three-dimensional software visualization. In: International Working
  Conference on Source Code Analysis and Manipulation (SCAM), IEEE, pp 163--174

\bibitem[{Storey et~al.(2005)Storey, {\v{C}}ubrani{\'c}, and
  German}]{storey2005use}
Storey MAD, {\v{C}}ubrani{\'c} D, German DM (2005) On the use of visualization
  to support awareness of human activities in software development: a survey
  and a framework. In: ACM Symposium on Software Visualization (SOFTVIS), ACM,
  pp 193--202

\bibitem[{Sun et~al.(2013)Sun, Wu, Liang, and Liu}]{sun2013survey}
Sun GD, Wu YC, Liang RH, Liu SX (2013) A survey of visual analytics techniques
  and applications: State-of-the-art research and future challenges. Journal of
  Computer Science and Technology 28(5):852--867

\bibitem[{Tang et~al.(2018)Tang, Rubab, Lai, Cui, Yu, and
  Wu}]{tang2018istoryline}
Tang T, Rubab S, Lai J, Cui W, Yu L, Wu Y (2018) {iStoryline}: Effective
  convergence to hand-drawn storylines. IEEE Transactions on Visualization and
  Computer Graphics

\bibitem[{Toosi et~al.(2018)Toosi, Son, and Buyya}]{toosi2018clouds}
Toosi AN, Son J, Buyya R (2018) {CLOUDS-Pi}: A low-cost {Raspberry-Pi} based
  micro data center for software-defined cloud computing. IEEE Cloud Computing
  5(5):81--91

\bibitem[{Toprak et~al.(2014)Toprak, Wichmann, and
  Schupp}]{toprak2014lightweight}
Toprak S, Wichmann A, Schupp S (2014) Lightweight structured visualization of
  assembler control flow based on regular expressions. In: IEEE Working
  Conference on Software Visualization (VISSOFT), IEEE, pp 97--106

\bibitem[{Trumper et~al.(2013)Trumper, D\"{o}llner, and
  Telea}]{trumper2013multiscale}
Trumper J, D\"{o}llner J, Telea A (2013) Multiscale visual comparison of
  execution traces. In: IEEE International Conference on Program Comprehension
  (ICPC), IEEE, pp 53--62

\bibitem[{Tymchuk et~al.(2016)Tymchuk, Merino, Ghafari, and
  Nierstrasz}]{tymchuk2016walls}
Tymchuk Y, Merino L, Ghafari M, Nierstrasz O (2016) Walls, pillars and beams: A
  {3D} decomposition of quality anomalies. In: IEEE Working Conference on
  Software Visualization (VISSOFT), IEEE, pp 126--135

\bibitem[{Ulan et~al.(2018)Ulan, H{\"o}nel, Martins, Ericsson, L{\"o}we,
  Wingkvist, and Kerren}]{ulan2018quality}
Ulan M, H{\"o}nel S, Martins RM, Ericsson M, L{\"o}we W, Wingkvist A, Kerren A
  (2018) Quality models inside out: Interactive visualization of software
  metrics by means of joint probabilities. In: IEEE Working Conference on
  Software Visualization (VISSOFT), IEEE, pp 65--75

\bibitem[{Urli et~al.(2015)Urli, Bergel, Blay-Fornarino, Collet, and
  Mosser}]{urli2015visual}
Urli S, Bergel A, Blay-Fornarino M, Collet P, Mosser S (2015) A visual support
  for decomposing complex feature models. In: IEEE Working Conference on
  Software Visualization (VISSOFT), IEEE, pp 76--85

\bibitem[{Vincur et~al.(2017)Vincur, Navrat, and Polasek}]{vincur2017vr}
Vincur J, Navrat P, Polasek I (2017) {VR City}: Software analysis in virtual
  reality environment. In: IEEE International Conference on Software Quality,
  Reliability and Security Companion (QRS-C)

\bibitem[{Wang et~al.(2019)Wang, Weatherston, Storey, and
  German}]{wang2019clonecompass}
Wang Y, Weatherston J, Storey MA, German D (2019) {CloneCompass}:
  Visualizations for exploring assembly code clone ecosystems. In: IEEE Working
  Conference on Software Visualization (VISSOFT), IEEE, pp 88--98

\bibitem[{Wilde and German(2018)}]{wilde2018merge}
Wilde E, German D (2018) {Merge-Tree}: Visualizing the integration of commits
  into linux. Journal of Software: Evolution and Process 30(2):e1936

\bibitem[{Wilhelm et~al.(2018)Wilhelm, Cakaric, Schuele, and
  Gerndt}]{wilhelm2018tool}
Wilhelm A, Cakaric F, Schuele T, Gerndt M (2018) Tool-based interactive
  software parallelization: a case study. In: IEEE/ACM International Conference
  on Software Engineering: Software Engineering in Practice Track (ICSE-SEIP),
  IEEE, pp 115--123

\bibitem[{Williams et~al.(2019)Williams, Bigelow, and
  Isaacs}]{williams2019visualizing}
Williams K, Bigelow A, Isaacs K (2019) Visualizing a moving target: A design
  study on task parallel programs in the presence of evolving data and
  concerns. IEEE Transactions on Visualization and Computer Graphics
  26(1):1118--1128

\bibitem[{Wu et~al.(2018)Wu, Chen, Sun, Xie, Cao, Liu, and
  Cui}]{wu2018streamexplorer}
Wu Y, Chen Z, Sun G, Xie X, Cao N, Liu S, Cui W (2018) Streamexplorer: a
  multi-stage system for visually exploring events in social streams. IEEE
  Transactions on Visualization and Computer Graphics 24(10):2758--2772

\bibitem[{Xu et~al.(2019)Xu, Wang, Yang, Wang, Qiao, Qin, Xu, Zhang, and
  Qu}]{xu2019clouddet}
Xu K, Wang Y, Yang L, Wang Y, Qiao B, Qin S, Xu Y, Zhang H, Qu H (2019)
  Clouddet: Interactive visual analysis of anomalous performances in cloud
  computing systems. IEEE transactions on visualization and computer graphics
  26(1):1107--1117

\bibitem[{Yi et~al.(2007)Yi, ah~Kang, and Stasko}]{yi2007toward}
Yi JS, ah~Kang Y, Stasko J (2007) Toward a deeper understanding of the role of
  interaction in information visualization. IEEE Transactions on Visualization
  and Computer Graphics 13(6):1224--1231

\bibitem[{Yoon et~al.(2013)Yoon, Myers, and Koo}]{yoon2013visualization}
Yoon Y, Myers BA, Koo S (2013) Visualization of fine-grained code change
  history. In: IEEE Symposium on Visual Languages and Human-Centric Computing
  (VL/HCC), IEEE, pp 119--126

\bibitem[{Zhao et~al.(2014)Zhao, Cao, Wen, Song, Lin, and
  Collins}]{zhao2014fluxflow}
Zhao J, Cao N, Wen Z, Song Y, Lin YR, Collins C (2014) {\#FluxFlow}: Visual
  analysis of anomalous information spreading on social media. IEEE
  Transactions on Visualization and Computer Graphics 20(12):1773--1782

\bibitem[{Zhu et~al.(2019)Zhu, Alderfer, Furqan, Nebolsky, Char, Smith,
  Villareale, and Onta{\~n}{\'o}n}]{zhu2019programming}
Zhu J, Alderfer K, Furqan A, Nebolsky J, Char B, Smith B, Villareale J,
  Onta{\~n}{\'o}n S (2019) Programming in game space: how to represent parallel
  programming concepts in an educational game. In: International Conference on
  the Foundations of Digital Games, pp 1--10

\bibitem[{{Zhu} et~al.(2019){Zhu}, {Nacenta}, {Akgun}, and
  {Nightingale}}]{zhu2019people}
{Zhu} X, {Nacenta} MA, {Akgun} O, {Nightingale} P (2019) How people visually
  represent discrete constraint problems. IEEE Transactions on Visualization
  and Computer Graphics pp 1--1, \doi{10.1109/TVCG.2019.2895085}

\bibitem[{Zirkelbach et~al.(2019)Zirkelbach, Krause, and
  Hasselbring}]{zirkelbach2019hands}
Zirkelbach C, Krause A, Hasselbring W (2019) Hands-on: experiencing software
  architecture in virtual reality. Tech. rep., Department of Computer Science,
  Kiel University, Germany

\end{thebibliography}

%\begin{thebibliography}{}
% etc
%\end{thebibliography}

\end{document}